\documentclass[10pt,twocolumn,conference,final]{IEEEtran}
\makeatletter
\def\ps@headings{%
\def\@oddhead{\mbox{}\scriptsize\rightmark \hfil \thepage}%
\def\@evenhead{\scriptsize\thepage \hfil \leftmark\mbox{}}%
\def\@oddfoot{}%
\def\@evenfoot{}}
\makeatother

\usepackage{cite}

\usepackage{epsfig} \usepackage{amsmath} \usepackage{amstext}
\usepackage{amsfonts} \usepackage{amssymb}
\usepackage{eucal} \usepackage{graphicx}
\usepackage{colordvi}
\usepackage{subfigure}
\usepackage{url}

\newcommand{\ls}[1] {\dimen0=\fontdimen6\the\font \lineskip=#1\dimen0

\advance\lineskip.5\fontdimen5\the\font \advance\lineskip-\dimen0
\lineskiplimit=.9\lineskip \baselineskip=\lineskip
\advance\baselineskip\dimen0 \normallineskip\lineskip
\normallineskiplimit\lineskiplimit \normalbaselineskip\baselineskip
\ignorespaces } \ls{1}

 \newtheorem{lma}{Lemma}[section]
\newtheorem{proposition}{Proposition}[section]
\newtheorem{thm}{Theorem}[section]

\newtheorem{corol}{Corollary}[section]
\newcommand{\Lt}{\tilde L}
\newcommand{\Zt}{\tilde Z}
\newcommand{\eat}[1]{}

\begin{document} %\baselineskip 0.1in

% paper title
% can use linebreaks \\ within to get better formatting as desired
\title{\LARGE Forward Correction and Fountain Codes in Delay Tolerant Networks}

% author names and affiliations
% use a multiple column layout for up to three different
% affiliations
\author{\IEEEauthorblockN{Eitan Altman}
\IEEEauthorblockA{INRIA, 2004
Route des Lucioles, \\ 
06902 Sophia-Antipolis Cedex, France\\
eitan.altman@sophia.inria.fr}
\and
\IEEEauthorblockN{Francesco De Pellegrini}
\IEEEauthorblockA{CREATE-NET, via Alla Cascata 56 c,\\
38100 Trento, Italy\\
francesco.depellegrini@create-net.org}
}

\date{} \maketitle

\begin{abstract}
Delay tolerant Ad-hoc Networks leverage the mobility of relay nodes to
compensate for lack of permanent connectivity and thus enable   
communication between nodes that are out of range of each other. To
decrease delivery delay, the information to be delivered 
is replicated in the network. Our objective in this paper is to study 
a class of replication mechanisms that include coding in order to improve the 
probability of successful delivery within a given time limit. We propose an 
analytical approach that allows to quantify tradeoffs between resources and 
performance measures (energy and delay). We study the effect of 
coding on the performance of the network while optimizing parameters that 
govern routing. Our results, based on fluid approximations, are compared
to simulations which validate the model.\footnote{A version of this paper appears in the proc. 
of IEEE INFOCOM 2009.}.
\end{abstract}

\thispagestyle{empty}

\section{Introduction}\label{sec:intro}

Delay tolerant Ad-hoc Networks make use of nodes' mobility to compensate for
lack of instantaneous connectivity. Information sent by a source to a
disconnected destination can be forwarded and relayed by other mobile nodes.
There has been a growing interest in such networks as they have the potential
of providing many popular distributed services \cite{burleigh_commag03,chaintreau07,spy_ton08}.

 A naive approach to forward a file to the destination is by epidemic routing in which
% cr 
%any mobile that has the message keeps on relaying it to any other
%mobile that enters its radio transmission range. 
any mobile that has the message keeps on relaying it to any other
mobile that falls within its radio range. 
% and of change
This would minimize the delivery delay  at the cost of inefficient use of network resources
(e.g. in terms of the energy used for flooding the network).
The need for more efficient use of network resources
%motivated the use of more economic forwarding such as the
% +-* I
motivated the use of less costly forwarding schemes such as the
two-hops routing protocols. In two-hops routing the source transmits copies of its 
message to all mobiles it encounters; relays transmit the message 
only if they come in contact with the destination. The two-hops protocol 
was originally introduced in \cite{tse_mobility02}.

% +-+ Francesco change below 
% In this paper we consider another aspect, i.e, the tradeoff between 
% network resources and delay, that of limited storage capacity at a node and/or
% finite contact durations. 
% We assume that the file to transfer is large with respect to the buffering
% capabilities of the nodes. As a result, it needs to be split
% into $K$ smaller units (which we call chunks or frames) and each
% such unit needs to be forwarded independently of the others;
% the message is considered to be well received only once all $K$ 
% frames are received at the destination. 
%
% The memory constraints due to buffering limitation make it advantageous
% to better organize the storage of information in the relay nodes.
% We aim at improving the efficiency of the DTN's operation by
% letting the source distribute not only the original frames but also
% additional redundant packets. This results in a
% spatial coding of the distributed storage of the frames.
% We consider coding based on either
% forward error correction techniques or on network coding approaches.
% Our main contribution is to provide close form expression for the
% performance of the DTNs (in terms of transfer delay and energy consumption)
% as a function of the coding that is used.

In this paper we consider another aspect, i.e, the tradeoff between 
network resources and delay. We assume that the file to be transferred 
needs to be split into $K$ smaller units: this happens due %can be due to the  
 to the finite duration of contacts between mobile nodes or when the 
file is large with respect to the buffering
 capabilities of nodes. Such $K$ smaller units (which we call chunks 
or frames) need to be forwarded independently of the others. The 
message is considered to be well received only if all $K$ frames 
are received at the destination. 

After fragmenting the message into smaller frames, it is convenient
to better organize the way information is stored in the relay nodes.
We aim at improving the efficiency of the DTN's operation by
letting the source distribute not only the original frames but also
additional redundant packets. This results in a spatial coding of 
the distributed storage of the frames. We consider coding based on either
forward error correction techniques or on network coding approaches.
Our main contribution is to provide a close form expression for the
performance of DTNs (in terms of transfer delay and energy consumption)
as a function of the coding that is used. Also, we derive scaling laws 
for the success probability of message delivery.

The paper is organized as follows. In Sec.~\ref{sec:related} we
revise the state of art and outline the major contributions of the
paper. In Sec.~\ref{sec:model} we describe the model of the system
and in Sec.~\ref{sec:redundancy} we derive the main results for the
case of erasure codes. Sec.~\ref{sec:fountain} is devoted to the
analysis of fountain codes; for both cases, we discover 
and then study an interesting phenomenon of phase transition. Sec.~\ref{sec:redundancy} 
and \ref{sec:fountain} involve also the design of energy-aware 
forwarding policies where the source forwards packets with some 
fixed probability that we optimize. In Sec.~\ref{sec:threshold} 
we study an alternative class of forwarding policies, namely 
 threshold type policies, that achieve the same energy restrictions. 
The performances of the aforementioned coding techniques 
in the case threshold policies are then derived. 
Sec.~\ref{sec:simulations} reports on simulation results 
in case of synthetic mobility and real-world traces. 
A concluding section ends this paper. \\

%Varios coding techniques had already been considered for improving the
%performance of DTNs. In \cite{fec1} ...

\noindent{\bf Related Works}\label{sec:related}
The idea to erasure code a message and distribute the generated code-blocks
over a large number of relays in DTNs has been addressed first in \cite{WJMK} 
and \cite{JDPF}. The technique is meant to increase the efficiency of DTNs 
under uncertain mobility patterns. In \cite{WJMK} the performance gain 
is compared to simple replication, i.e. the technique of releasing additional 
copies of the same message. The benefit of erasure coding is quantified in that 
work via extensive simulations for various routing protocols, including 
two-hops routing.%  Also, under the hypothesis that message delays are i.i.d. 
% Pareto r.v., the authors show that the expected delay converges to a finite value.

In \cite{JDPF}, the case of non-uniform encounter patterns is addressed, showing
that there is strong dependence of the optimal successful delivery probability
on the allocation of replicas over different paths. The authors evaluate several 
allocation techniques; also, the problem is proved to be NP--hard.

General network coding techniques \cite{FBW} have been proposed for DTNs. In \cite{LLL}
ODE based models are proposed under epidemic routing. Semi-analytical numerical
results are reported describing the effect of finite buffers and contact
times; a prioritization algorithm is also proposed.

The work in \cite{WB} addresses the use of network coding techniques for
stateless routing protocols under intermittent end-to-end connectivity.
 A forwarding algorithm based on network coding is specified, showing a 
clear advantage over plain probabilistic routing in the delivery of multiple packets.

Finally, an architecture supporting random linear coding in challenged
wireless networks is reported in \cite{FSCB}. \\

\noindent{\bf Novel contributions}

The main contribution of this paper is the closed form description of the
performance of Delay tolerant Ad-hoc Networks under the two-hops relaying 
protocol when a message is split into multiple frames. Our fluid model 
accounts both for the overhead of the forwarding mechanism, captured in the form
of a given bound on energy, and the probability of successful delivery
of the entire message to the destination within a certain deadline. The 
effect of coding is included in the model and both erasure codes and 
fountain codes are accounted for in closed form. The two coding strategies are 
characterized in the case of static probabilistic forwarding policies and in 
the case of threshold policies.

Leveraging the model, the asymptotic properties of the system are derived
 in the form of scaling laws. In particular, there exists a threshold 
law ruling the success probability which ties together the main parameters
of the system. 

To the best of the authors' knowledge, the results contained in this work
represent the first description in closed form of the behavior of
erasure codes and fountain codes in challenged networks. 

%%%%%%%%%%%%%%%%%%%%%%%%%%%%%%%%%%%%%%%%%%%%%%%%%%%%%%%%%%%%%%%%%%%%%%%%%%%%%%%%%%%%%%%%%%%%%%%%%%%%%%%%%%%%%%%%%%%%%%%%
%%%%%%%%%%%%%%%%%%%%%%%%%%%%%%%%%%%%%%%%%%%%%%%%%%%%%%%%%%%%

\section{The model}\label{sec:model}

%%%%%%%%%%%%%%%%%%%%%%%%%%%%%%%%%%%%%%%%%%%%%%%%%%%%%%%%%%%%
%%%%%%%%%%%%%%%%%%%%%%%%%%%%%%%%%%%%%%%%%%%%%%%%%%%%%%%%%%%%%%%%%%%%%%%%%%%%%%%%%%%%%%%%%%%%%%%%%%%%%%%%%%%%%%%%%%%%%%%%
For the ease of reading, the main symbols used in the paper are reported in Tab.~\ref{tab:notation}.
\begin{table}
\centering
\begin{tabular}{|p{0.10\columnwidth}|p{0.8\columnwidth}|}
\hline
{\it Symbol} & {\it Meaning}\\
\hline
$N$ & number of nodes (excluding the destination)\\
$K$ & number of frames composing the message\\
$M$ & number of frames needed to decode with success probability $1-\delta$, $\delta>0$ (fountain codes)\\
$H$ & number of redundant frames\\
$\lambda$ & inter-meeting intensity\\
$\tau$ & timeout value\\
$X_i(t)$ & number of nodes having frame $i$ at time $t$ (excluding the destination)\\
$X(t)$ & sum of the $X_i$s\\
$\overline X(t)$ & sum of the $X_i$s when $u_i(t)=1,\, \forall i=1,2,\ldots,K$ \\
${ \cal E}(t)$& energy expenditure by the whole network in $[0,\,t\,]$\\
$x$ & maximum number of copies due to energy constraint \\
$z$ & :=$X(0)$ \\ 
$\varepsilon$ & energy per frame\\
$u_i(t)$ & forwarding policy for frame $i$ \\ 
$p_i$ & static forwarding policy for frame $i$; $\mathbf p=(p_1,p_2,\ldots,p_K)$ \\ 
$p$ & sum of the $p_i$s \\ 
$D_i(\tau)$ & probability of successful delivery of frame $i$ by time $\tau$\\
$P_s(\tau)$ & probability of successful delivery of the message by time $\tau$; $P_s(\tau,K,H)$ is used to stress the dependence on $K$ and $H$\\
\hline
\end{tabular}\caption{Main notation used throughout the paper}\label{tab:notation}
\end{table}

Consider a network that contains $N+1$ mobile nodes. We assume that two nodes are able to
communicate when they are within reciprocal radio range, and communications are
bidirectional. We also assume that contact intervals are sufficient to exchange all
frames: this let us consider nodes {\em meeting times} only, i.e., time instants
at which a pair of not connected nodes fall within reciprocal radio range.

Also, let the time between contacts of pairs of nodes be exponentially distributed 
with given inter-meeting intensity $\lambda$. The validity of this 
model been discussed in \cite{GNK}, and its accuracy has been shown for a number 
of mobility models (Random Walker, Random Direction, Random Waypoint).
%+-+ Added by Francesco in order to justify exponential assumptions
\footnote{We recall that studies based on traces collected from real-life mobility~\cite{chaintreau07} 
argue that inter-contact times may follow a power-law distribution, but recently the 
authors of~\cite{karagiannis07} have shown that these traces and 
many others exhibit exponential tails after a cutoff point.} 

% Furthermore, the regime we consider is that of sparse networks 
% where inter-meeting intensities are very small on the scale of message 
% lifetime. When this is the case, point processes satisfy Poisson 
% assumptions due to the rare events law \cite{where?}.

We assume that the transmitted message is relevant during some time
$\tau$. We do not assume any feedback that allows the source or other
mobiles to know whether the messages has made it successfully
to the destination within time $\tau$.

The source has a message that contains $K$ frames.
If at time $t$ it encounters a mobile which does not have
any frame, it gives it frame $i$ with probability
$u_i$, and we let $u=\sum_i u_i \leq 1$ (we shall consider both the case where $u_i$
depends on $t$ and the case where it does not).
For the message to be relevant, all $K$ frames
should arrive at the destination by time $\tau$. 
Let $X_i (t)$ be the number of the mobile nodes (excluding the 
destination) that have at time $t$ a copy of frame $i$. 
Denote by $D_i(\tau)$ the probability of a successful delivery of frame $i$
by time $\tau$. Then, given the process $X_i$ (for which a fluid approximation 
will be used), we have
\begin{eqnarray}
\nonumber
%\overline D_i(t) & := & 1 - D_i (t) = P(t_i > \tau )  \\ &=&
D_i(\tau) & = 1 - \exp \left( - \lambda \int_0^\tau X_i(s) ds \right)
\label{dl1}
\end{eqnarray}
This expression has been derived in \cite{ABD} for the fluid model.

The probability of a successful delivery of the message by time $\tau$ is thus
\[
P_s (\tau) = \prod_{i=1}^K D_i( \tau )
= \prod_{i=1}^K
\left[ 1 - \exp \left( - \lambda \int_0^\tau X_i(s) ds \right) \right]
\]
%+cr+ some comments on the docoupling assumption
where we assumed that the success probability of a given frame is independent 
of the success probability of other frames; this decoupling assumption is 
confirmed by our numerical experiments.
% and cr

%%%%%%%%%%%%%%%%%%%%%%%%%%%%%%%%%%%%%%%%%%%%%%%%%%%%%%%%%%%%
\subsection{Fluid Approximations}
%%%%%%%%%%%%%%%%%%%%%%%%%%%%%%%%%%%%%%%%%%%%%%%%%%%%%%%%%%%%

Let $X(t)=\sum_{i=1}^K X_i(t)$. 
Then we introduce the following standard fluid approximation (based on mean field analysis) \cite{ZNKT}
\begin{equation}
\frac{ d X_i (t) } { dt} = u_i (t) \lambda (N - X(t))
\label{dyn1}
\end{equation}
%+-+ Moved before
%where $u_i (t) \in [ u_{\min} , 1 ]$, $u_{\min}>0$.
Taking the sum over all $i$, we obtain the separable differential equation 
\begin{equation}
\frac{ d X (t) } { dt} = u (t) \lambda (N - X(t))
\label{dyn2}
\end{equation}
whose solution is
% \[
% X(t) = N + (z-N) e^{ - \lambda \int_s^t u(v) dv }
% \]
\[
X(t) = N + (z-N) e^{ - \lambda \int_0^t u(v) dv }, \quad X(0)= z
\]

Thus, $X_i(t)$ is given by the solution of
\begin{equation}
\frac{d X_i (t) }{dt} = - u_i(t) \lambda
(z-N) e^{ - \lambda \int_0^t u(v) dv }
\end{equation}

%%%%%%%%%%%%%%%%%%%%%%%%%%%%%%%%%%%%%%%%%%%%%%%%%%%%%%%%%%%%
\subsection*{Constant policies}
%%%%%%%%%%%%%%%%%%%%%%%%%%%%%%%%%%%%%%%%%%%%%%%%%%%%%%%%%%%%

In the case of constant policies, we let $u_i(t)=p_i$, $\mathbf p:=(p_1,p_2,\ldots,p_K)$, 
and $p=\sum_i p_i$. Hence it follows
\begin{eqnarray}
X_i(t)=X_i(0)+(N-z)\frac{p_i}{p} \left [ 1-e^{-\lambda pt} \right ]
\end{eqnarray}

% Let us assume, $s=0$ and $X_i(0)=0$ for $\forall i=1,2,\ldots,K$. Hence,
% \begin{eqnarray}
% X_i(t)=N\frac{p_i}{p} \Big ( 1-e^{-\lambda pt} \Big )
% \label{xit}
% \end{eqnarray}
Let us assume $X_i(0)=0$ for $\forall i=1,2,\ldots,K$: hence,
\begin{eqnarray}
X_i(t)=N\frac{p_i}{p} \Big ( 1-e^{-\lambda pt} \Big )
\label{xit}
\end{eqnarray}

%%%%%%%%%%%%%%%%%%%%%%%%%%%%%%%%%%%%%%%%%%%%%%%%%%%%%%%%%%%%
\subsection{Taking Erasures into Account}
%%%%%%%%%%%%%%%%%%%%%%%%%%%%%%%%%%%%%%%%%%%%%%%%%%%%%%%%%%%%

So far we have assumed that the transmission of a frame is always
successful. Assume that this is not the case and that the
transmission of a frame fails with some probability $q$. We assume
that the process describing whether packets transmissions are  
successful or not is i.i.d. %
%+cr+
%We assume moreover that a packet that
%suffers from unrecoverable transmission errors at a mobile is
%discarded and does not occupy memory space in the relay node; at the
%next time that the mobile meets another mobile that has a packet to
%transmit to it, the transmitter can relay the packet.
%Losses such as those just described
%do not need an extra modeling: we may replace
%the rate  $\lambda$ of inter-meetings between two nodes
% by the rate $\lambda ( 1 - q ) $ of the potentially
%successful inter-meetings of the nodes. This can be used in
%describing in the equations we derived for the dynamics of the
%system and for its performance measures. 
We assume moreover that a packet that
suffers from unrecoverable transmission errors at a mobile is
discarded so that it does not occupy memory space in the relay node; 
this ensures that such a mobile node can still act as a relay at 
the next meeting with another mobile having a packet to be transmitted.

Losses such as those just described
do not need an extra modeling: we may replace
the rate  $\lambda$ of inter-meetings between two nodes
 by $\lambda ( 1 - q )$, i.e., the rate of the potentially
successful inter-meetings of the nodes. This can be used in the 
equations that we derived in describing the dynamics of the system 
and its performance measures.

% +cr+ changed 
%An additional type of loss may occur at the destination: in case
%it is not mobile and is connected to the an external network
%(possibly a wired one), then losses may occur in that part of the
%network. Assume that a loss there occurs with probability $q'$.
%We do not assume any feedback that would allow the
%DTN to know about events that occur at the  external network.
%In order to be able to recover from such losses we assume
%that the destination may keep receiving copies of the same
%frame. In particular, a mobile that has transmitted a
%frame to the destination will keep the copy and could try
%to retransmit it to the destination at future inter-meeting
%occasions. This additional type of loss process does not
%alter the fluid dynamics of $X_i(t)$. Its impact on the performance
%is by replacing $\lambda$ in (\ref{dl1}) by $\lambda(1-q')$.
An additional type of loss may occur at the destination: this is 
the case when it is not mobile and it is connected to an external 
network (possibly a wired one): in this case losses may occur 
in that part of the network. Assume that a loss there occurs 
with probability $q'$. We do not assume any feedback that would 
allow the DTN to know about events that occur at the external network.
In order to be able to recover from such losses we assume
that the destination may keep receiving copies of the same
frame. In particular, a mobile that has transmitted a
frame to the destination will keep the copy and could try
to retransmit it to the destination at future inter-meeting
occasions. This additional type of loss process does not
alter the fluid dynamics of $X_i(t)$. Its impact on the performance
is by replacing $\lambda$ in (\ref{dl1}) by $\lambda(1-q')$.
\bigskip

%%%%%%%%%%%%%%%%%%%%%%%%%%%%%%%%%%%%%%%%%%%%%%%%%%%%%%%%%%%%
\subsection{\bf Non-constrained problem.}
%%%%%%%%%%%%%%%%%%%%%%%%%%%%%%%%%%%%%%%%%%%%%%%%%%%%%%%%%%%%

The success probability when using ${\bf p } $ is
\begin{eqnarray}
\lefteqn{
\nonumber
P_s(\tau , {\bf p } )=
\prod_{i=1}^{K}\Big (1-\exp(-\lambda \int_0^\tau X_i(v)dv) \Big )  } \\
\nonumber
&=&\prod_{i=1}^{K}
\Big[ 1-\exp \Big (- \frac{ \lambda }{ p } \int_0^\tau N{p_i}
\Big ( 1-e^{-\lambda p v} \Big ) dv \Big ) \Big] \nonumber \\ 
& = & \prod_{i=1}^{K} Z (p_i)
\end{eqnarray}
\begin{equation}
\label{Zi}
\mbox{ where }
Z(p_i) := 1-\exp\Big( L(\tau, p) p_i \Big)
\mbox{ and }
\end{equation}
\[
L( \tau , p ):= \frac {N}{p^2 }  \Big ( 1- \lambda p \tau -
e^{-\lambda p \tau} \Big )
\]

For fixed ratios $p_i/p$,
$P_s(\tau , {\bf p } )$ is increasing in $p$ and
is maximized at $p=1$.

Let $P_s^*(\tau)$ be the optimal delivery probability
for the problem of maximizing $P_s (\tau)$ with $p$ fixed.
The proof of the following can be found in the Appendix.% \cite{report}.
\begin{thm}\label{thm:maximize}
${\bf p^*}=(p/K,...,p/K)$ is the unique solution to
the problem of maximizing $P_s(\tau)$ s.t.
$ \sum_i p_i = p $, $p_i \geq 0 $.
\label{thm3p1}
\end{thm}

%%%%%%%%%%%%%%%%%%%%%%%%%%%%%%%%%%%%%%%%%%%%%%%%%%%%%%%%%%%%%%%%%%%%%%
\subsection{Constrained problem.}
%%%%%%%%%%%%%%%%%%%%%%%%%%%%%%%%%%%%%%%%%%%%%%%%%%%%%%%%%%%%%%%%%%%%%%

Denote by $ {\cal E} (t) $ the
energy consumed by the whole network
for the transmission of the message during the time interval $[0,t]$.
It is proportional to $X(t) - X(0)$ since we assume
that the message is transmitted only to mobiles that
do not have the message, and thus the number of transmissions
of the message during $[0,t]$ plus the number of mobiles that had it at time zero
equals to the number of mobiles that have it. % Added by Francesco %
Also, let $\varepsilon>0$ be the energy spent to forward a frame
during a contact. We thus have $ { \cal E}(t) = \varepsilon (X(t) - X(0))$.
In the following we will denote $x$ as the maximum number of copies that can be released 
due to energy constraint.
% We are interested in the optimal probability of successful delivery of the
% message by time  $\tau$ under the constraint that the energy consumption till time $\tau$ is upper bounded
% by some constant ${\cal E}$.

We compute in particular the optimal probability of successful delivery of the
message by some time  $\tau$ %for these forwarding schemes
under the constraint that the energy consumption till time $\tau$ is bounded
by some positive constant.% ${\cal E}>0$. 

Define $\overline X(t)$ to be the solution of (\ref{dyn2}) when
$u(t)=1$, i.e.
\[
\overline X(t) = N + (z-N) e^{ - \lambda t }
\]
Note that $X(t)= \overline X(pt )$.

Denote $ \sigma(z) := \overline X^{-1}(x+z) $ given $\overline X(0)=z$,
which is the time elapsed until  $x$ extra nodes (in addition to
the initial $z$ ones)  receive the message in the uncontrolled
system. We have
\begin{equation}
\sigma (z) = -\frac{1}{\lambda} \log \left(\frac{ N-x-z }{ N-z  } \right)
\label{sigma}
\end{equation}

For $p=\sigma(z)/\tau$ we obtain the expression
%\[
%L(\tau,p) =
%\frac{ N}{ p^2 } \Big( 1 + \log \big(
%\frac{ N - z- x }{ N - z } \Big)
%- \frac{ N - z- x }{ N - z } \Big)
%\]
\begin{equation}
\label{ltp}
L(\tau,p) =
\frac{ N}{ p^2 } \Big( \log \Big(
1 - \frac{  x }{ N - z } \Big)
+ \frac{ x }{ N - z } \Big) 
\end{equation}

\begin{thm}
\label{res1}
Consider the problem of maximizing $P_s(\tau)$ subject to
a constraint on the energy $ {\cal E} (\tau)\leq \varepsilon x $.
\\
(i) If $\overline X (\tau) \leq x+z $
(or equivalently, $\tau \leq \sigma (z) $),
then a control policy $u$ is optimal if and only if
$p_i=1/K$ for all $i$.
\\
(ii) If $\overline X ( u_{\min} \tau ) >  z+ x  $ (or equivalently,
$ u_{\min} \tau >  \sigma( z ) $), then there
is no feasible control strategy.
\\
(iii) If $\overline X (\tau) > z+ x > \overline X ( u_{\min} \tau ) $
(or equivalently, $\tau > \sigma(z) > u_{\min} \tau $),
then the best control policy is given by $p_i = p^*/K$ 
where
\begin{equation}
\label{Pstar}
p^* = \frac{\sigma(z) }{\tau}
\end{equation}
and the optimal value is
% \[
% P_s^*( \tau ) = \Big( 1 - \exp ( L ( p^* , \tau ) p^* /K ) \Big)^K
% \]
\[
P_s^*( \tau ) = \Big[ 1 - \Big( 1 -  \frac{  x }{ N - z } \Big)^{\frac N{p^* K}}\exp \Big(-\frac{ N}{ p^* K } \frac{ x }{ N - z }  \Big)\Big ]^K
\]
\end{thm}

\begin{IEEEproof}
%{\bf Proof.}
Part (i) and  (ii) are obvious. Part (iii) follows from the fact that $X\big (\frac{\sigma(z)}{\tau}\tau\big )=\overline X(\sigma(z))=x+z$, so that the energy bound is attained for $p^*=\frac{\sigma(z) }{\tau}$; also, the expression for $P_s^*( \tau )$ follows from an immediate application of the result in Thm.~\ref{thm:maximize} to (\ref{Zi}).
% \begin{eqnarray}
% Z(p_i)&&=1-\exp\left( L( \tau , p^* ) \frac{ p^* }{K} \right)\nonumber\\
%       &&=1 - \Big( 1 -  \frac{  x }{ N - z } \Big)^{N/(p^* K)}\exp \Big(\frac{ N}{ p^* K } \frac{ x }{ N - z } \Big)\nonumber
% \end{eqnarray}
%\endpf
\end{IEEEproof}

% revised till here
%%%%%%%%%%%%%%%%%%%%%%%%%%%%%%%%%%%%%%%%%%%%%%%%%%%%%%%%%%%%%%%%%%%%%%%%%%%%%%%%%%%%%%%%%%%%%%%%%%%%%%%%%%%%%%%%%%%%%%%%%%%%%%%%%%
%%%%%%%%%%%%%%%%%%%%%%%%%%%%%%%%%%%%%%%%%%%%%%%%%%%%%%%%%%%%%%%%%%%%%%

\section{Adding Fixed amount of Redundancy}\label{sec:redundancy}

%%%%%%%%%%%%%%%%%%%%%%%%%%%%%%%%%%%%%%%%%%%%%%%%%%%%%%%%%%%%%%%%%%%%%%
%%%%%%%%%%%%%%%%%%%%%%%%%%%%%%%%%%%%%%%%%%%%%%%%%%%%%%%%%%%%%%%%%%%%%%%%%%%%%%%%%%%%%%%%%%%%%%%%%%%%%%%%%%%%%%%%%%%%%%%%%%%%%%%%%%
\begin{figure*}[t]
\centering 
\subfigure{\includegraphics [width=5.5cm] {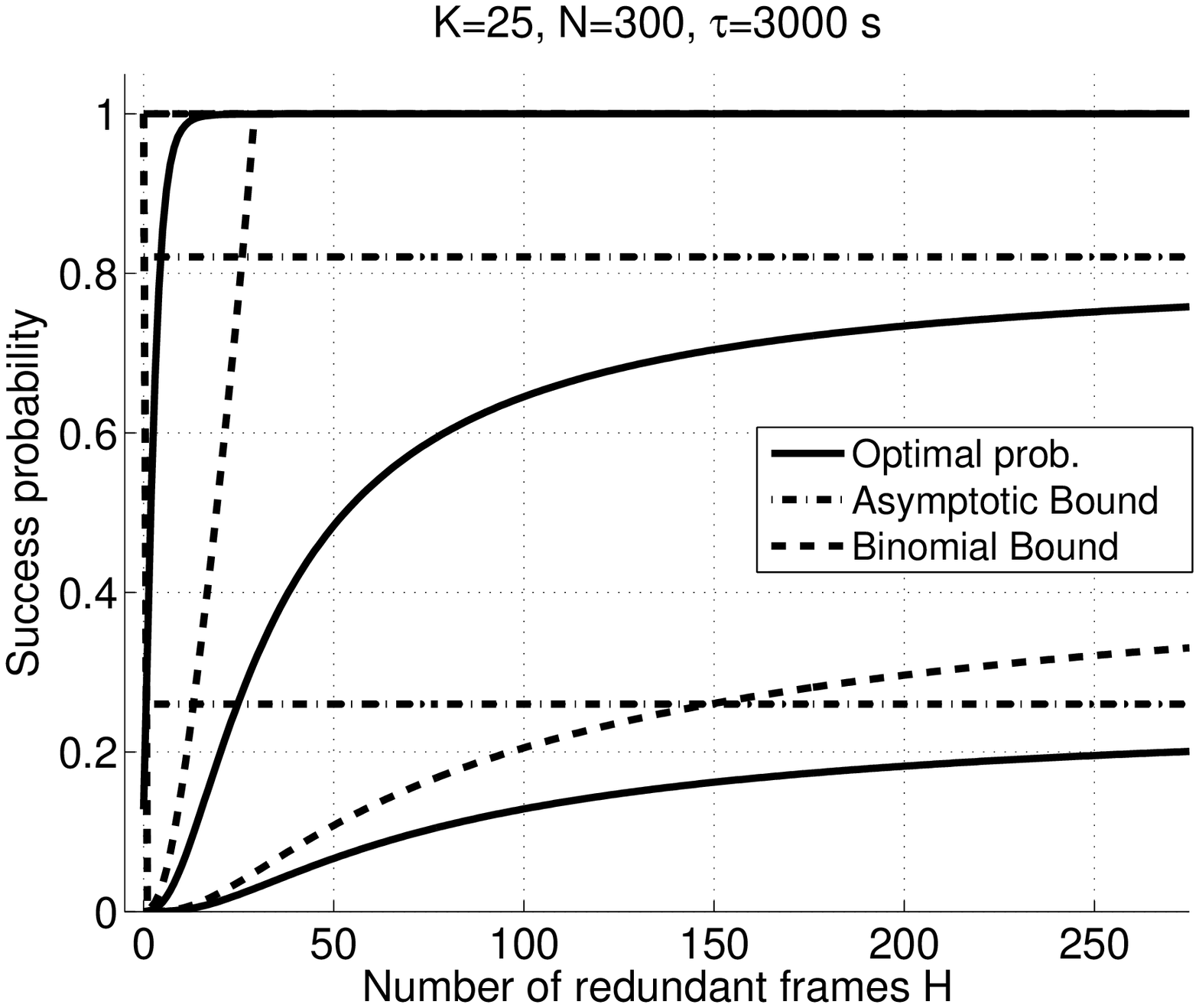}
{\scriptsize 
\put(-10,110){$\lambda_1$}
\put(-10,83){$\lambda_2$}
\put(-10,27){$\lambda_3$}}}\hskip4mm
\subfigure{\includegraphics [width=5.5cm] {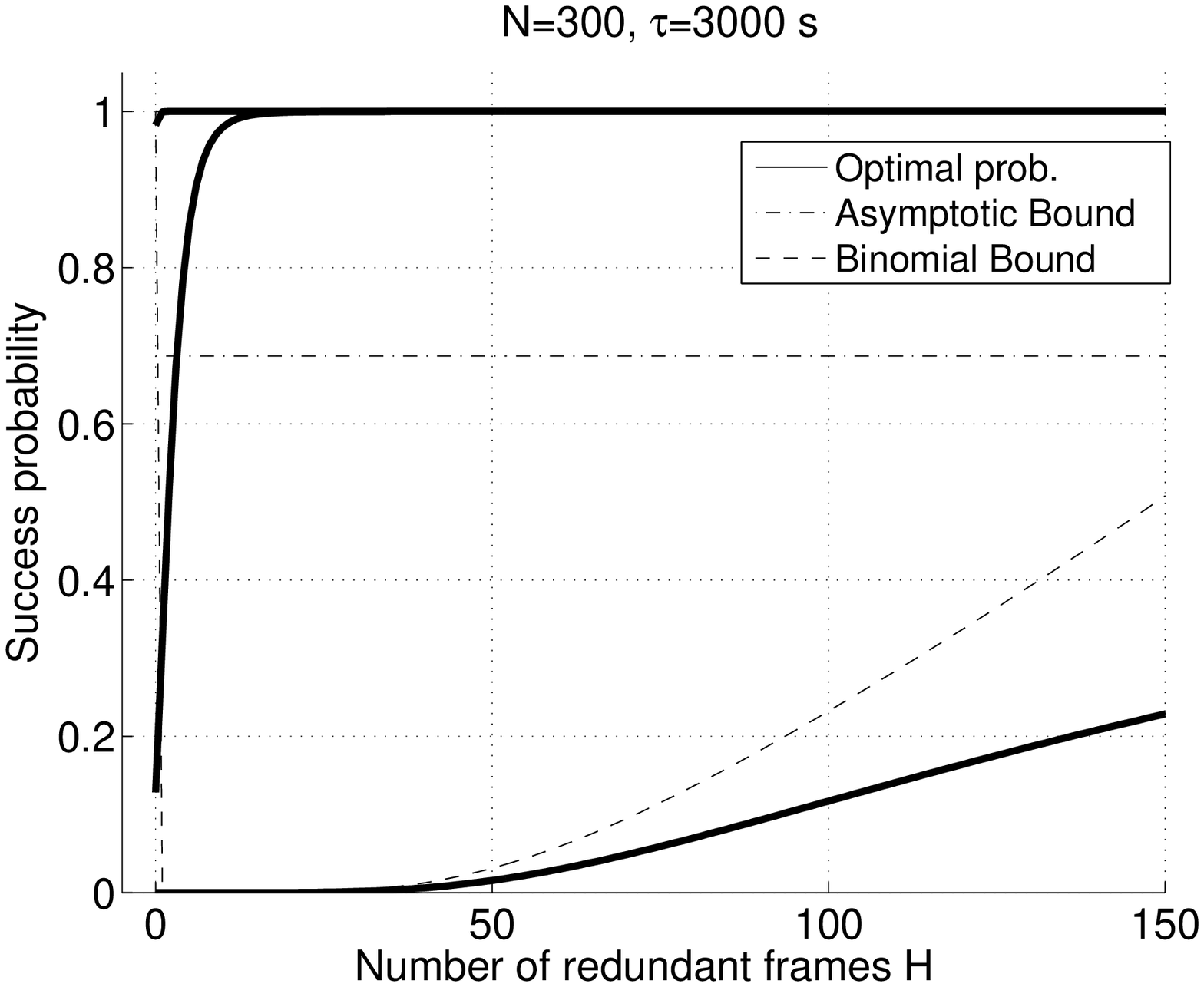}
{\scriptsize \put(-136,118){$K=10$}\put(-130,90){$K=25$}\put(-30,40){$K=60$}}}\hskip4mm
\subfigure{\includegraphics [width=5.5cm] {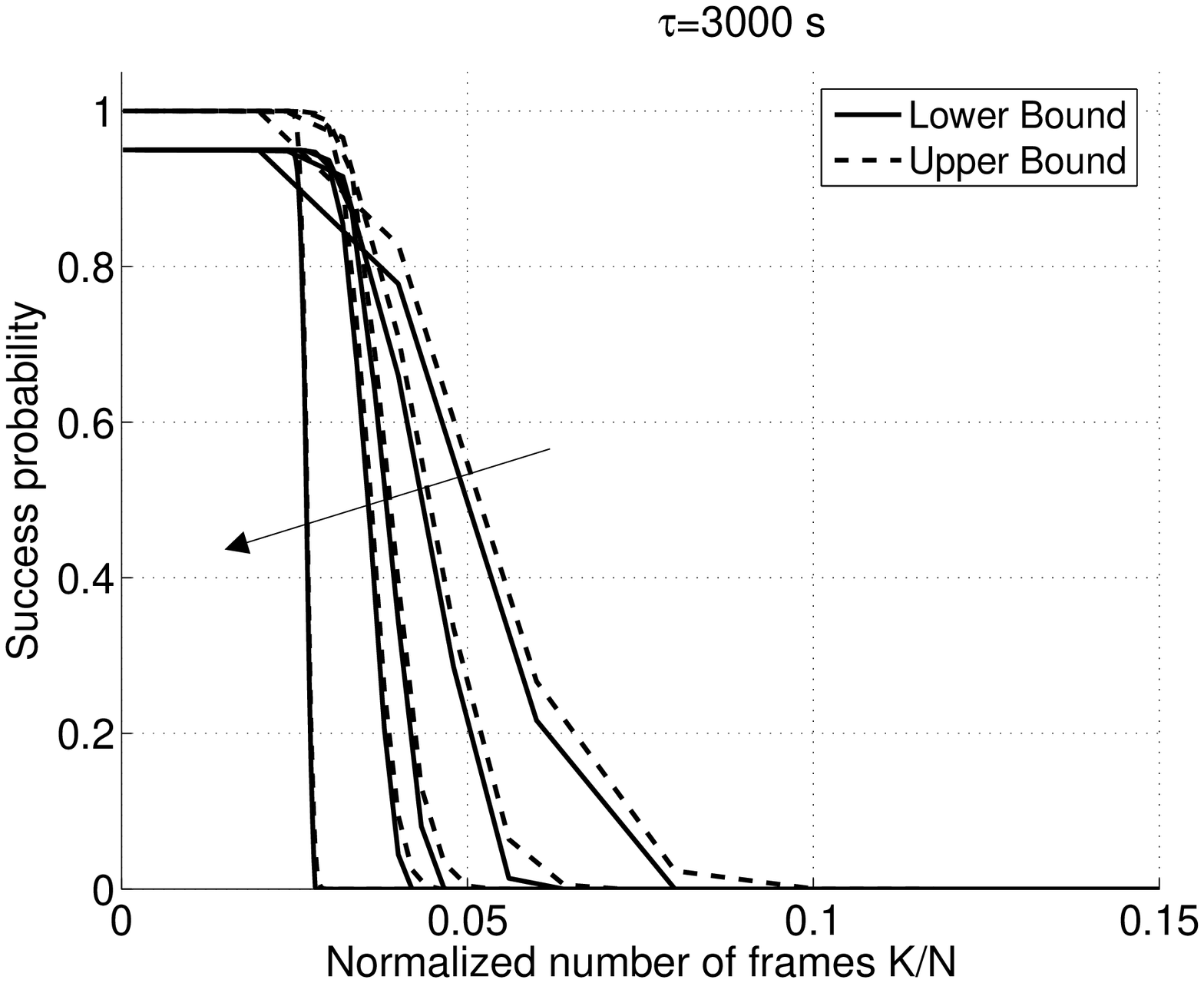}\put(-90,73){{\scriptsize $N=50,125,300,500,5000$}}}
\put(-495,110){a)}\put(-325,110){b)}\put(-157,110){c)}
\caption{a) Success probability for $N=300$, $K=25$, $p=1$ under different values of $\lambda$; $\lambda_1=0.22 \cdot 10^{-03} s^{-1}$, $\lambda_2=0.091 \cdot 10^{-03} s^{-1}$ $\lambda_3=0.065 \cdot 10^{-03} s^{-1}$ b) Success probability for 
$\lambda_1$, $N=300$, $p=1$ under different values of $K$ c) Upper and lower bound $p=1$, $\delta=0.02$ as given in Cor.~\ref{cor:fountain_bound}, for various values of $N=50,125,300,500,5000$, versus normalized number of transmitted frames $K/N$.}\label{fig:fig1}
\end{figure*}

We add $H$ redundant frames and consider
the new message that now contains $K+H$ frames.
If at time $t$ the source encounters a mobile which does not have
any frame, it gives it frame $i$ with probability
$p$.

Let $S_{n,p}$ be a binomially distributed r.v. with parameters $n$ and $p$,
i.e.
\[
P(S_{n,p}=m) = B(p,n,m) := { n \choose m } p^m (1-p)^{n-m} 
\]
The probability of a successful delivery of the message
by time $\tau$ is thus
\[
P_s (\tau,K,H) = \sum_{j=K}^{K+H} B( D_i(\tau) , K+H , j  ), 
\]
where $D_i(\tau) = 1 - \exp ( - \lambda \int_0^\tau X_i(s) ds)$.

\subsection{Main Result}

\begin{lma}
The maximum of $P_s(\tau,K,H)$ over $\{ p_m \geq 0, m=1,...,K+H\}$
under the constraint $\sum_{i=1}^{K+H} p_i = p $ is achieved at
$p_m = p/(K+H)$.
\label{equal}
\end{lma}
The proof is delayed to the Appendix.% \cite{report}.

Using the same arguments as those that led to Theorem \ref{res1},
together with Lemma \ref{equal} yields the following:

\begin{thm}\label{thm:main_theorem}
Consider the problem of maximizing $P_s(\tau,K,H)$ over
$p_i, i=1,...,K+H,$ subject to
a constraint on the energy $ {\cal E} (\tau)\leq \varepsilon x $.
\\
(i) If $\overline X (\tau) \leq x+z $
(or equivalently, $\tau \leq \sigma (z) $),
then a control policy $u$ is optimal if and only if
% -->F--> %%%%%%%%%%%%%%%%%%%%%%%%%
%$p_i=1/K$ for all $i$.
$p_i=1/(K+H)$ for all $i$.
%%%%%%%%%%%%%%%%%%%%%%%%%%%%%%%%%%
\\
(ii) If $\overline X ( u_{\min} \tau ) >  z+ x  $ (or equivalently,
$ u_{\min} \tau >  \sigma( z ) $), then there
is no feasible control strategy.
\\
(iii) If $\overline X (\tau) > z+ x > \overline X ( u_{\min} \tau ) $
(or equivalently, $\tau > \sigma(z) > u_{\min} \tau $),
then the best control policy is given by $p_i = p^*/(K+H)$ where
$ p^* $ is given in (\ref{Pstar}) and the optimal value is
\[
P_s^*( \tau , K, H ) = \sum_{j=K}^{K+H} B \left( \widehat p , K+H , j \right)
\]
%\mbox{where  }
where
\[
\widehat p = 1 - \exp ( L ( p^* , \tau ) p^* /(K+H) )
\]
\end{thm}

%%%%%%%%%%%%%%%%%%%%%%%%%%%%%%%%%%%%%%%%%%%%%%%%%%%%%%%%%%%%%%%%%%%%%%
\subsection{Properties and Approximations}
%%%%%%%%%%%%%%%%%%%%%%%%%%%%%%%%%%%%%%%%%%%%%%%%%%%%%%%%%%%%%%%%%%%%%%

We now derive further characterizations for the optimal 
success probability; these results will 
provide both bounds for the case of block coding and help also 
in the analysis of fountain codes. 

\begin{corol}\label{cor:cormon}
$P_s^* (\tau,K,H)$ is increasing with $H$.
\end{corol}

%\noindent{\bf Proof.}
\begin{IEEEproof}
We make the following observation.
Fix $K$ and $H$ and a vector $( p_1 , ... , p_{K+H} )$ whose
entries sum up to $p^*$ (given in (\ref{Pstar})).
Then the success probability is the same as
when increasing the redundancy to
$H+1$ and using the vector $( p_1 , ... , p_{K+H} , 0  )$.
By definition,
the latter is strictly smaller than $P_s^* ( \tau , K , H + 1 )$
(which is the optimal success probability with $H+1$ redundant
packets).

The above is in particular true when taking
$p_i=p^*/(K+H)$ (which maximizes $P_s(\tau, K, H)$),
and hence
$P_s^* ( \tau, K, H ) < P_s^* ( \tau , K , H+1 ) $.
\end{IEEEproof}
%\endpf

%%%%%%%%%%%%%%%%%%%%%%%%%%%%%%%%%%%%      Inserted by Francesco   %%%%%%%%%%%%%%%%%%%%%%%%%%%%%%%%%%%%%%%%%%

Now we provide two useful bounds for $P_s^*(\tau,H,K)$. 

First, it is possible to derive the asymptotic approximation of $P_s^*(\tau,H,K)$ for $H\rightarrow \infty$. For the sake of notation,
let $V=L(p,\tau)p$; the following holds
\begin{eqnarray}
&&P_s^*(\tau,H,K)= e^V \sum_{s=K}^{H+K}{{H+K}\choose s} \Big ( e^{-\frac{V}{H+K}} - 1\Big )^s \nonumber \\
&&                =1-e^V \sum_{s=0}^{K-1}{{H+K}\choose s} \Big  ( \frac{-V}{(H+K)^s} + o((H+K)^{-s})\Big )\nonumber
\end{eqnarray}\label{eq:asympt}
For large values of $H+K$, the $s$-th term of the right-hand summation writes
\begin{eqnarray}
&&{{H+K}\choose s} \Big ( \frac{-V}{H+K}\Big )^s = \frac{(H+K)!}{s!(H+K-s)!} \frac{(-V)^s}{(H+K)^s}\nonumber \\
&&\sim \frac{(-V)^s}{s!e^s} \frac{1}{\Big( 1-\frac{s}{H+K} \Big )^{H+K}} \sim \frac{(-V)^s}{s!}\nonumber
\end{eqnarray}\label{eq:sthterm}
where the Stirling formula applies, $n!\sim \sqrt{2\pi n} \; (\frac{n}{e})^n$, with $f \sim g$
meaning $\lim_{H\rightarrow \infty} f/g=1$.
\begin{corol}\label{cor:stirling_bound}
For $K\geq 1$ and $\tau\geq 0$,
\[
\lim_{H\rightarrow \infty} P_s^*(\tau,H,K)= 1-e^{L(p,\tau)p} \sum_{s=0}^{K-1}\frac{[-L(p,\tau))p]^s}{s!}
\]
\end{corol}
We note that in sight of the monotonicity stated in Cor.~\ref{cor:cormon}, the limit is reached from below.

%%%%%%%%%%%%%%%%%%%%%%%%%%%%%%%%%%%%      End    %%%%%%%%%%%%%%%%%%%%%%%%%%%%%%%%%%%%%%%%%%

Also, the binomial bound \cite[Thm 1.1]{Bollobas} applies. %
% -->F--> %%%%%%%%%%%%%%%%%%%%%%%%%
% For $1 \leq m \leq n-1 $, define $u$ through $m= \lceil upn \rceil $.
For $1 \leq m \leq n-1 $, and $u>1$, define $u$ through $m= \lceil upn \rceil $.
%%%%%%%%%%%%%%%%%%%%%%%%%%%%%%%%%%%%%%%%%%%%%%%%%%%%%%%%%%%%%%%%%%%%%%%%%%%%%%%%%%%%%%%%%
Then
% \begin{equation}
% P(S_{n,p} \geq m ) \leq
% \label{bela}
% \end{equation}
% \[
% % -->F--> %%%%%%%%%%%%%%%%%%%%%%%%%
% %\frac{1}{\sqrt{2 \pi }}\frac{u}{1-u}
% \frac{1}{\sqrt{2 \pi }}\frac{u}{1-u}
% %%%%%%%%%%%%%%%%%%%%%%%%%%%%%%%%%%%
% \left( \frac{n}{m(n-m)} \right)^{1/2} u^{-upn } \left( \frac{1-p}{1-up }
% \right)^{(1-up)n}
% \]
\begin{equation}\label{bela}
P(S_{n,p} \geq m ) \leq \frac{n^{1/2}}{(2 \pi m(n-m))^{1/2}}  \frac{u^{1-upn}}{1-u}          \left( \frac{1-p}{1-up }\right)^{(1-up)n}\nonumber
\end{equation}
%%%%%%%%%%%%%%%%%%%%%%%%%%%%%%%%%%%%      Inserted by Francesco   %%%%%%%%%%%%%%%%%%%%%%%%%%%%%%%%%%%%%%%%%%

In Fig~\ref{fig:fig1}a) we reported the numerical comparison of the optimal delivery probability as a function of 
the number of redundant frames, for a particular setting. The asymptotic bound is reported as an horizontal dot-slashed 
line; the binomial bound is reported with slashed line. We notice that the binomial bound proves a good approximation 
for lower success probabilities, i.e., at smaller values of $\lambda$. %: in such cases, in fact,
%$S_{n,p}$ has a larger tail. 
The relative increase of the success probability under erasure codes is apparent: for instance, the success 
probability with $\lambda=0.22 \cdot 10^{-03} s^{-1}$ increases from $0.12$ for $H=0$ to one with a few redundant 
frames ($H=12$); the maximum attainable improvement, though, is dictated by the upper bound.

In Fig~\ref{fig:fig1}b) the optimal delivery probability is depicted as a function of the number of redundant frames, 
at the increase of $K$; even in this case the binomial bound proves a better approximation for lower success probabilities, 
as it appears in the case of $K=60$. 

%%%%%%%%%%%%%%%%%%%%%%%%%%%%%%%%%%%%%%%%%%%%%%%%%%%%%%%%%%%%%%%%%%%%%%%%%%%%%%%%%%%%%%%%%%
\subsection{Phase transition }
%%%%%%%%%%%%%%%%%%%%%%%%%%%%%%%%%%%%%%%%%%%%%%%%%%%%%%%%%%%%%%%%%%%%%%%%%%%%%%%%%%%%%%%%%%

% +cr+ commented this in more depth as requested by the secodn reviewer
% In what follows we elaborate based on result from Thm.~\ref{thm:main_theorem}, and we study the case 
%of large values of $N$ under the asymptotic regime $H+K=(H+K)(N)\leq N$. 
In what follows we elaborate based on result from Thm.~\ref{thm:main_theorem}, and we study the case 
of large values of $N$. Let us assume that the total number of frames, $H+K$, grows as a function of $N$,
i.e.,  $H+K=(H+K)(N)$.
 
The question we would like to answer is, in the asymptotic regime, what is the effect
of redundancy onto the delivery probability, that is, how the number of redundant 
frames should scale with respect to the total number of frames.

% cr end of change

We assume throughout that the following limits exist:
\[
\widehat K := \lim_{N\rightarrow \infty} K(N)/N \mbox{ and }
\widehat H := \lim_{N\rightarrow \infty}H(N)/N 
\]
Notice that this implies of course that the energy constraint should 
grow at most linearly with $N$ and we thus assume that the limit 
$\widehat x := \lim_{N\rightarrow \infty} x(N)/N>0$ exists \footnote{In what follows
we will exclude the trivial case $\widehat x=0$, which corresponds to the case 
when no relaying is allowed.}. 
\begin{proposition}
\label{phase1}
Introduce the threshold
\[
\Gamma_0 := \lambda\tau \Big( 1+\frac{\widehat{x}}{\log(1-\widehat{x})} \Big),
\]
then, the following holds:
\\
{\small
\hspace{-2mm}
\fbox{
$
\begin{array}{c}
\lim_{N\rightarrow \infty} P_s(\tau,K,H)= 
\begin{cases}
0 & \mbox{if} \ \widehat K+\widehat H  \leq \Gamma_0\\[2mm]
0 & \displaystyle
 \mbox{if} \  \widehat K+\widehat H > \Gamma_0
   \ \mbox{and} \ \widehat K >\Gamma_0\\[2mm]
1 & \mbox{if} \
\displaystyle
\widehat K+\widehat H>\Gamma_0 \ 
   \mbox{and} \ \widehat K < \Gamma_0
\end{cases}
\end{array}
$
}
}

\end{proposition}
The proof is in the Appendix. We conclude that there exist a phase-transition effect. Its threshold $\Gamma_0$ is the same 
as that we shall obtain later for the fountain codes.

The way we interpret such result is that, in order to deliver with high probability the message, for large $N$, 
the $K$ message frames should not exceed such threshold, but the sum of the encoded ones should.

%%%%%%%%%%%%%%%%%%%%%%%%%%%%%%%%%%%%      End    %%%%%%%%%%%%%%%%%%%%%%%%%%%%%%%%%%%%%%%%%%

\section{Fountain Codes}\label{sec:fountain}

Each time the source meets a node, it sends to it (with probability $p$)
a packet obtained by generating a  new random linear combination
of the  $K$ original packets. Using fountain codes, we know that
for any $\delta$
in order for the destination to be able to decode the original message
with probability at least $1-\delta$, it has to receive
at least $M := K\log(K/\delta)$ packets \cite[Chap 50]{McKay}.
%+-+ just to explain that we use it for the asymptotics
A useful expression for large $K$ will be used later: 
if we write $M=K(1+\alpha)$ then $\alpha $ that guarantees
that the destination can decode the original message with
probability of at least $1-\delta$ is given by
\[
\alpha = \frac{ ( \log (K/\delta ) )^2 }{ \sqrt{K} }
\]
The number of packets that reach the destination during the time
interval $[0,\tau]$ has a Poisson distribution with parameter
% -->F--> %%%%%%%%%%%%%%%%%%%%%%%%%
$-L( \tau , p^* )p^* $
%%%%%%%%%%%%%%%%%%%%%%%%%%%%%%%%%%
where $p^*$ is given in (\ref{Pstar}) and where
$L( \tau , p^* ) $  is given in (\ref{ltp}).

The probability that less than $M$ packets reach the destination is
given by
\begin{equation}\label{eq:PM}
P_M ( \tau ) = \sum_{i=0}^{M-1}
\frac{ (-L( \tau , p^* )p^*)^i }{i! } \exp( L (\tau , p^* )p^*  )
\end{equation}
We conclude that
\(
P_s^*( \tau ) \geq 1 - \delta - P_M ( \tau ) .
\)

%%%%%%%%%%%%%%%%%%%%%%%%%%%%%%%%%%%%      Inserted by Francesco   %%%%%%%%%%%%%%%%%%%%%%%%%%%%%%%%%%%%%%%%%%

Finally, we can leverage Cor. \ref{cor:stirling_bound} and obtain
\begin{corol}\label{cor:fountain_bound}
Given $p^*$ as in (\ref{Pstar}),
\[
1 - \delta - P_M ( \tau ) \leq P_s^*( \tau ) \leq 1 - P_M ( \tau )
\]
\end{corol}
We notice that the bound on the right-end of the inequality is obtained 
by using redundancy as in the previous section for $H\geq M$, then  taking the 
limit of $H\rightarrow \infty$. 

\subsection{Numerical examples: a phase transition}

In Fig~\ref{fig:fig1}c) we reported the representation of the bounds described above: the two bounds provide a
tight characterization of the performances of fountain codes; in particular, we notice that $P_M(\tau)$, which in
fact is the CDF of a Poisson r.v., tends to one as $K$ increases: this causes the success probability to tend to zero as
$K$ increases. The intuition is that, for a large number of transmitted frames, the probability of receiving all of 
them successfully within the given deadline decreases faster than the gain obtained by adding redundancy through fountain codes.

But, the numerical insight of the model says more of the performance attained by fountain codes. In fact, in Fig~\ref{fig:fig1}c) 
we reported $P_s^*$ versus the  normalized number of transmitted frames $K/N$, for increasing $N$. Interestingly, 
the success probability becomes more and more close to a step function at the increase of $N$. We thus observe a phase 
transition: above a given number of transmitted frames, the success probability vanishes, and below the same threshold 
success occurs with probability one.

We shall study this phenomenon analytically in the next subsection.

\subsection{Analysis Asymptotic behavior}

We now want to understand the behavior of the fountain codes for large values of $N$ in the 
asymptotic regime $K=K(N)\leq N$; in what follows, we will 
assume that $\widehat K := \lim_{N\rightarrow \infty} K(N)/N$ exists. 
This implies of course that the energy constraint should grow linearly
with $N$ and we thus assume that the limit 
$\widehat x := \lim_{N\rightarrow \infty} x(N)/N$ exists. 
We can rewrite $-L( \tau , p^* )p^*=N\cdot \Gamma_0^{(0)} $, where we obtain
\[
\Gamma_0^{(N)}=-\frac {1}{p^*}  \Big ( 1- \lambda p^* \tau -
e^{-\lambda p^* \tau} \Big )=\lambda\tau 
\Big( 1+\frac{\frac{x}{N-z}}{\log(1-\frac{x}{N-z})} \Big)
\]
\[
\Gamma_0 := \lim_{N \to \infty} \Gamma_0^{(N)} = \lambda\tau 
\Big( 1+\frac{\widehat{x}}{\log(1-\widehat{x})} \Big)
\]
We have
\[
P_M^{(N)} ( \tau ) = \sum_{i=0}^{K(N)(1+\alpha)-1}
\frac{(N\cdot \Gamma_0^{(N)})^i}{i!} \exp(-N\cdot \Gamma_0^{(N)})
\]
We notice that $P_M^{(N)} (\tau)=
Pr\{X_N\leq K(N)(1+\alpha)-1\}$, where $X_N$ 
is a Poisson r.v. From the Strong Law of Large Numbers
we have $ \lim_{N \to \infty} {X_N}/{N} = \Gamma_0 $ P-a.s.
Thus
\[
\lim_{N \to \infty} P_M^{(N)} (\tau ) =
1-\lim_{N \to \infty} Pr( {X_N} \geq K(N) )
\]
\[
=\begin{cases}
1, & \mbox{if} \quad \widehat K > \Gamma_0 \\
 0, & \mbox{if} \quad \widehat K < \Gamma_0 
\end{cases}
\]
from which we can deduce 

\fbox{
$
\lim_{N\rightarrow \infty} P_s(\tau)
\begin{cases}
\geq 1-\delta, & \mbox{if} \quad \widehat K < \Gamma_0\\
= 0, & \mbox{if} \quad \widehat K > \Gamma_0
\end{cases}
$
}

where we used Cor.~\ref{cor:fountain_bound}.

This demonstrates the effect of phase transition 
observed numerically in Fig~\ref{fig:fig1}c).

%%%%%%%%%%%%%%%%%%%%%%%%%%%%%%%%%%%%              End             %%%%%%%%%%%%%%%%%%%%%%%%%%%%%%%%%%%%%%%%%%

\section{Threshold policies}\label{sec:threshold}

\begin{figure*}[t]
\centering 
\subfigure{\includegraphics [width=5.5cm] {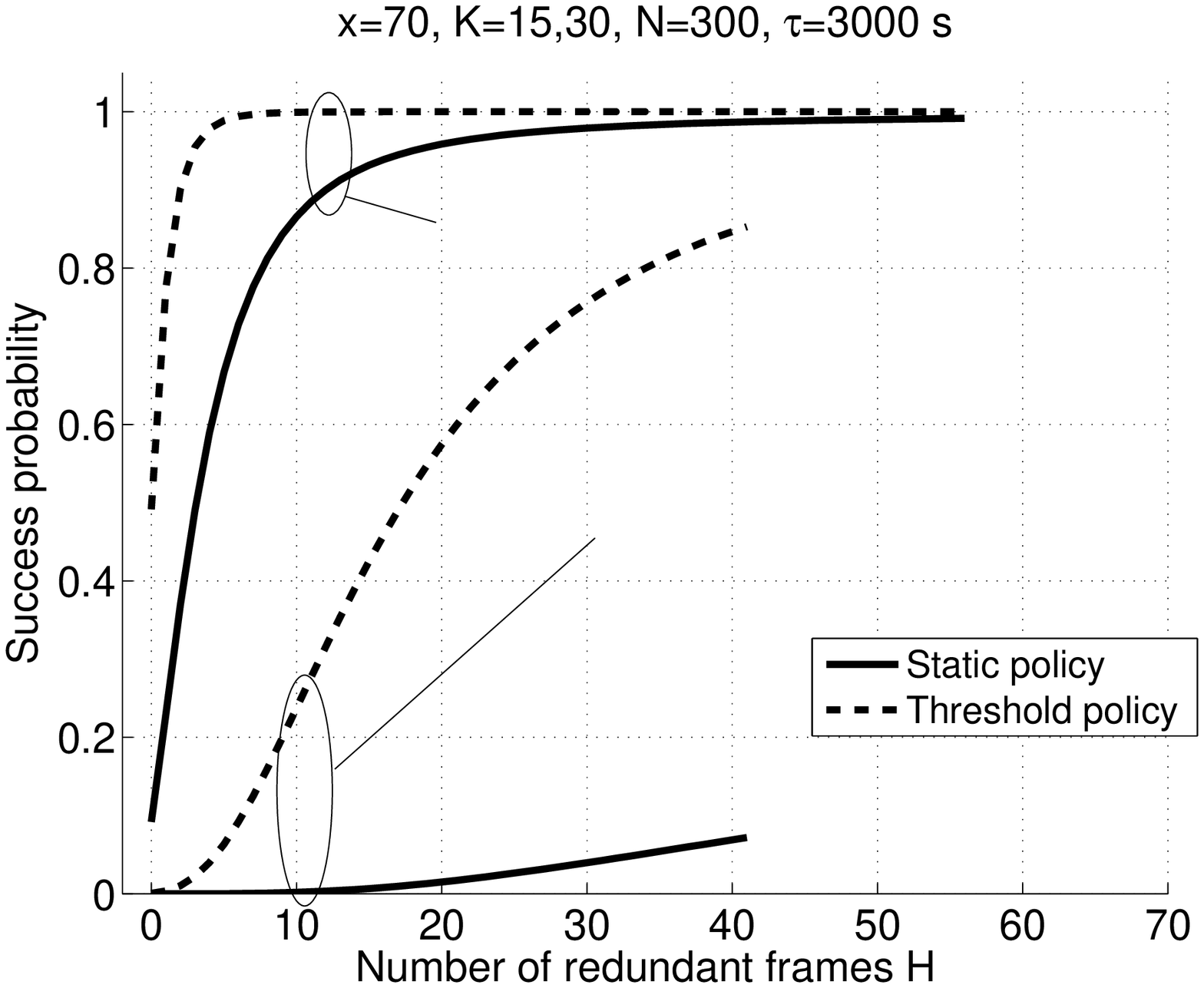}
{\small \put(-97,99){$K=15$}
\put(-78,60){$K=30$}}
}\hskip10mm%\hskip4mm
\subfigure{\includegraphics [width=5.5cm] {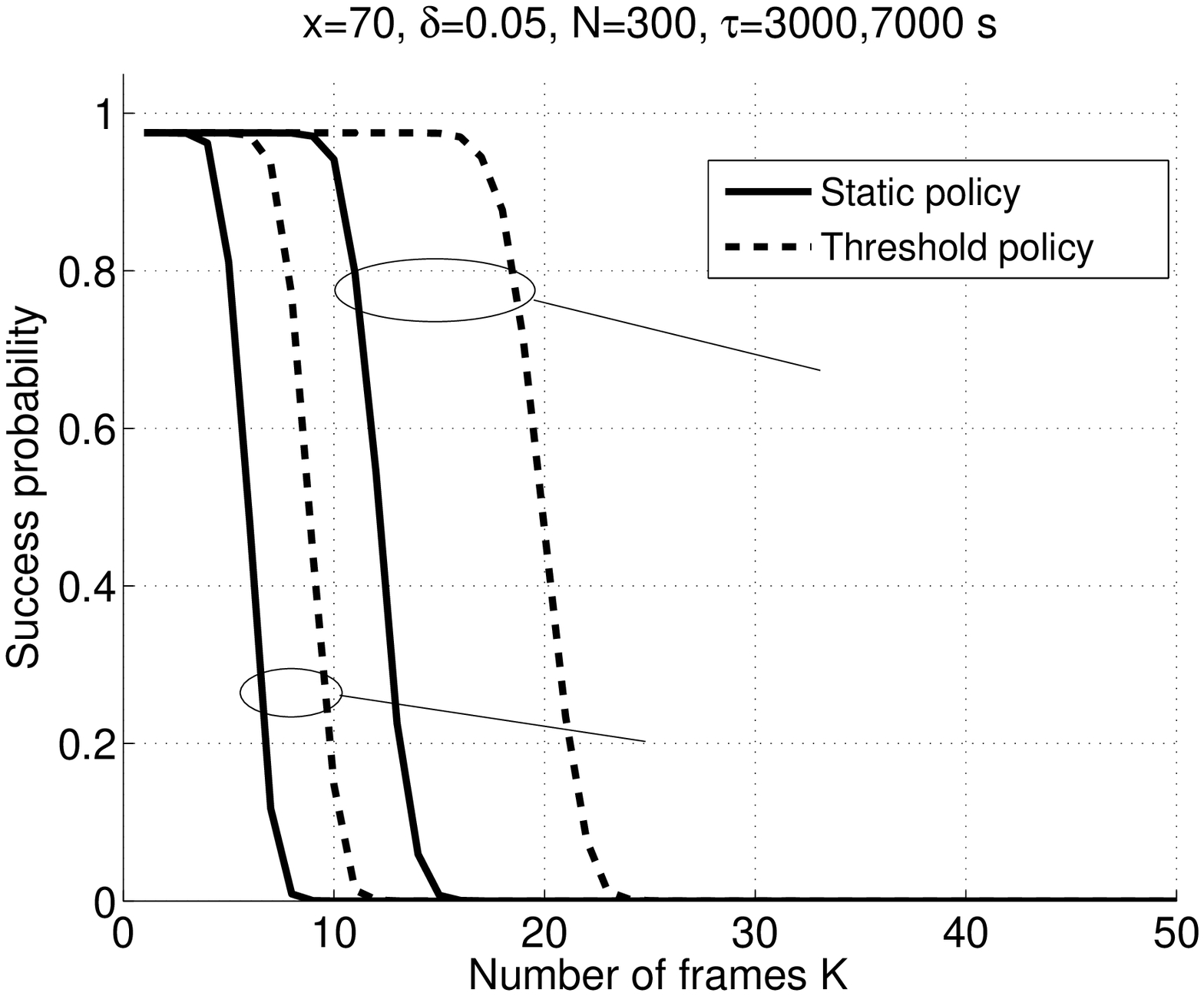}
{\small \put(-48,80){$\tau=7000 s$}
\put(-70,30){$\tau=3000 s$}}
}%\hskip4mm
\put(-350,110){a)}\put(-162,110){b)}
\caption{A comparison of success probability with erasure codes when using threshold and static policies; $N=300$, $x=70$, $\tau=3000,7000$: a) Erasure Codes, $K=15,30$ b) Fountain codes, $\delta=0.05$.}\label{fig:fig2}
\end{figure*}

The way the energy constraints are handled so far is
by using only a fraction of the transmission opportunities.
This was done uniformly in time by transmitting at a
probability that is time independent; we will refer to this strategy 
as {\em static} policy. In this section we consider
an alternative way to distribute the transmissions:
we use every possible transmission opportunity till some
time limit and then stop transmitting.
This is motivated by the "spray and wait" policy \cite{SPR} that
is known to trade off very efficiently message delay and
number of replicas in case of a single message.

In addition we need to specify the values of $p_i$: the
probability packet $i$ when there is an opportunity to transmit
(and the time limit has not yet elapsed). We have $\sum_{i=1}^K p_i = 1$.

In the lack of redundancy, we recall that the success probability writes % is given by
\[
P_s(\tau , {\bf p } )=
\prod_{i=1}^{K}\Big (1-\exp(-\lambda \int_0^\tau X_i(v)dv) \Big ) 
\]

Clearly, transmitting always when there is an opportunity to transmit
is optimal if by doing so the energy constraint are not violated.
This can be considered to be a trivial time limit policy with
a limit of $r=\tau$. Otherwise, the optimal value
$r$ for a threshold is the one that achieves the energy constraints,
or equivalently, the one for which the expected number of transmissions
during time interval $[0,r]$ by the source is $x$; as before, $x$ is related 
to the total constraint on the energy through the constraint 
$ {\cal E} (\tau)\leq \varepsilon x $.
The value $r$ of the time limit is then
given by $\sigma(z)$, see eq. (\ref{sigma}).

Since we considered a time limit policy, $X_i(v)$ first grows
till the time limit $r$ is reached and then it stays unchanged
during the interval $(r,\tau]$. For the non-trivial case where
$r = \sigma(z)$ we thus have:

\[
P_s(\tau , {\bf p } )  = 
\prod_{i=1}^{K}
\Big[ 1-\exp \Big (- { \lambda } \int_0^{\sigma(z)} N{p_i}
\Big ( 1-e^{-\lambda v}
\Big )  dv
\]
\begin{equation}
- \lambda ( \tau - \sigma(z)) X_i ( \sigma(z))
\Big ) \Big] \nonumber = \prod_{i=1}^{K} \Zt (p_i)
\label{eq:thr_ps}
\end{equation}
where $ \Zt(p_i) := 1-\exp\Big( p_i \Lt(\tau)  \Big) ,$
$ p_i \Lt( \tau  ):= -\lambda \,\int_0^{\tau} X_i(v)dv $
and where we have by eq. (\ref{xit}) and (\ref{sigma})
\[
X_i(\sigma(z))=N p_i \Big ( 1-e^{-\lambda \sigma(z)} \Big )= \frac{N p_i x}{N-z}
\]
Also, the integral can be expressed as
%\begin{equation}\label{eq:integral}
\[
\int_0^{\sigma(z)} X_i(v)dv=-\frac{Np_i}{\lambda} \left [ - \frac{x}{N-z}+\log \Big( 1 - \frac{x}{N-z} \Big )\right ]
% \nonumber
%&&=\frac{Np_i}{\lambda} \left [ \sum_{n=2}^{\infty} \frac{1}{n}\Big( \frac{-x}{N-z} \Big )^n\right ]\nonumber=\frac{Np_i}{\lambda}\frac{x}{N-z}\log \Big(1-\frac{x}{N-z}\Big )
\]
%\end{eqnarray}
%where we notice that the series expansion for $\log(1+v)$ has been used for $|v|=|x/(N-z)|<1$

In particular, the following holds
{\small
\[
\Lt( \tau  )= \frac{Nx}{N-z}+N\log
\Big (1-\frac{x}{N-z} \Big )- \lambda
(\tau - \sigma(z))\frac{Nx}{N-z}
\]
\[
=-\frac{Nx}{N-z}\lambda\tau - \beta(z)
\mbox{, \qquad where}
\]
}
\[
\beta(z)=-\frac{Nx}{N-z} -N\Big (1 - \frac{x}{N-z}\Big )\log\Big (1 - \frac{x}{N-z}\Big )\geq0.
\]

Finally, the $p_i$'s are selected to be all equal (and to sum to 1)
due to the same arguments as in the proof Theorem \ref{thm3p1} as here too,
$\log (\tilde Z)$ is concave in its argument.

In the case of redundancy, the calculations are similar to those in eq. (\ref{eq:thr_ps}), where
{\small
\begin{eqnarray}\label{eq:thr_ps_red}
P_s(\tau,H,K)=\sum_{s=K}^{H+K}{{H+K}\choose s} \tilde Z^s (p_i)(1-\tilde Z(p_i))^{N-s}
\end{eqnarray}
}
Hence,  given $H\geq 0$, it holds
\begin{eqnarray}\label{eq:thr_all}
\lefteqn{ P_s^*(\tau,H,K) =
\exp\left( -\frac{\frac{Nx}{N-z}\lambda\tau+\beta(z)}{H+K} \right) } \\
& \times & \sum_{s=K}^{H+K}{{H+K}\choose s} \left [ \exp\left(
\frac{\frac{Nx}{N-z}\lambda\tau+\beta(z)}{H+K} \right)
- 1 \right ]^s\nonumber
\end{eqnarray}
Here again, we used the fact that
the success probability is maximized for equal $p_i$'s, i.e.
\[
p_i^*(t)=
\begin{cases}
1/(H+K)& \quad{\mbox{if}}\quad t\leq \sigma(z)\\
0&\quad{\mbox{if}}\quad t > \sigma(z)
\end{cases}
\]
and $P_s^*(\tau,H,K)$ is given as in eq. (\ref{eq:thr_all}).
The proof follows the same lines as that of Lemma \ref{equal}.

As a final remark, consider the trivial case where the energy constraint is not active: $\tau < \sigma(z)$, i.e., it is optimal 
to transmit all the $x$ packets up to time $\tau$; in this case results from Thm.~\ref{thm:main_theorem} hold.

Now we want to specialize the use of threshold policies in the case of fountain codes when the energy constraint is active: 
$\tau \geq \sigma(z)$. In this case, the number of packets that reach the destination during the time interval $[0,\tau]$ 
has a Poisson distribution with parameter $\Lambda=-\Lt( \tau )=\frac{Nx}{N-z}\lambda\tau+\beta(z)$; the probability that 
less than $M=K\log(K)$ packets reach the destination is given by
% \[
% P_M ( \tau ) = \exp\left( {-\frac{Nx}{N-z}\lambda\tau - \beta(z)}
% \right)
% \]
% \[
% \times \sum_{i=0}^{M-1}
% \frac1{i!} \Big (\frac{Nx}{N-z}\lambda\tau +\beta(z) \Big )^i
% \]
\[
P_M ( \tau ) = \exp\left( {-\frac{Nx}{N-z}\lambda\tau - \beta(z)}\right) %
\sum_{i=0}^{M-1}\frac1{i!} \Big (\frac{Nx}{N-z}\lambda\tau +\beta(z) \Big )^i
\]

and the statement of Cor.~\ref{cor:fountain_bound} holds accordingly (notice that when the energy constraint is not active, 
we fall back to the original form of Cor~\ref{cor:fountain_bound}).
\begin{figure*}[t]
\centering 
\subfigure{\includegraphics [width=5cm] {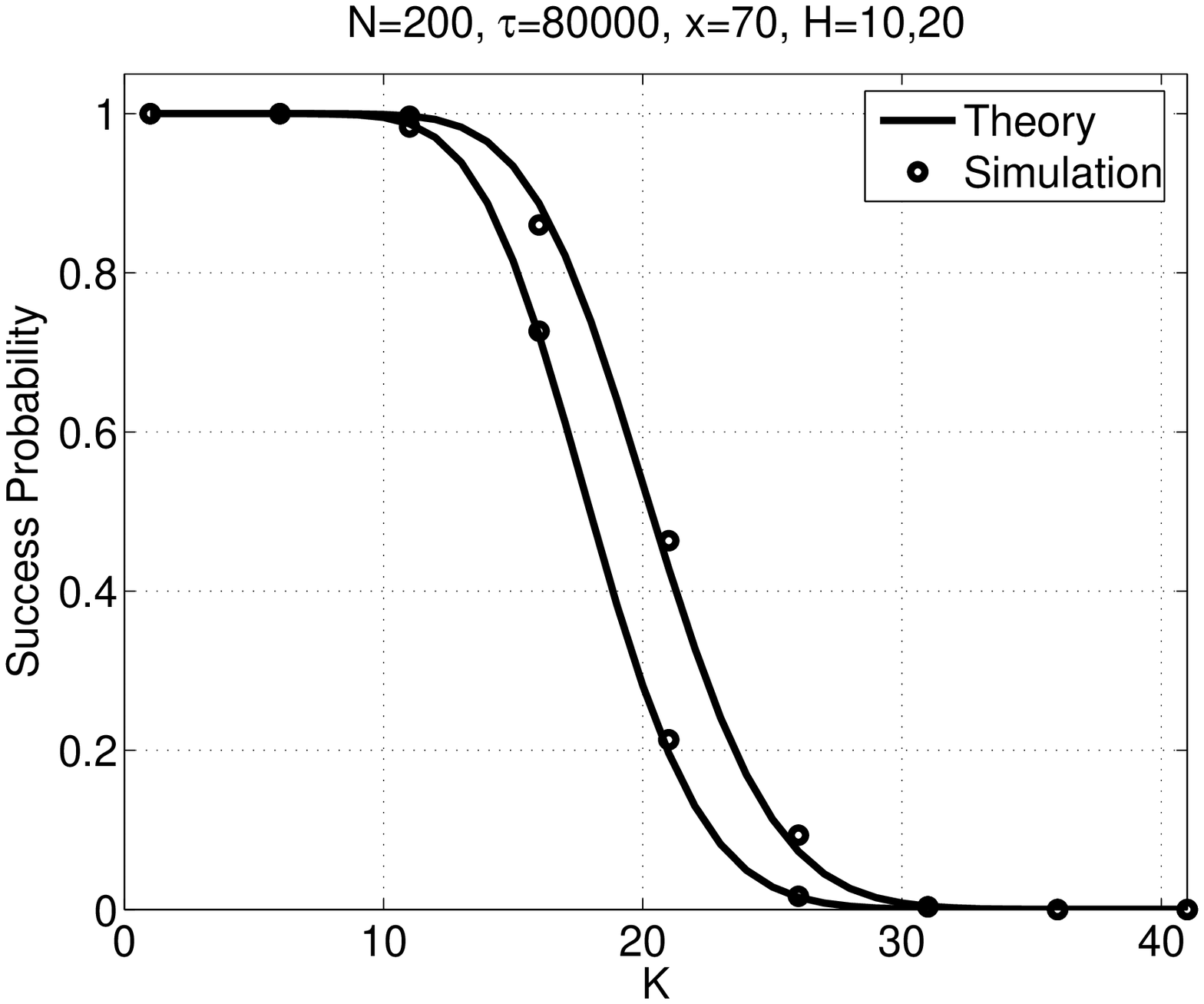}
\put(-40,50){\small $H=20$}\put(-43,47){\vector(-1,-1){10}}
\put(-120,35){\small $H=10$}\put(-85,45){\vector(1,1){10}}}\hskip5mm%\hskip4mm
\subfigure{\includegraphics [width=5cm] {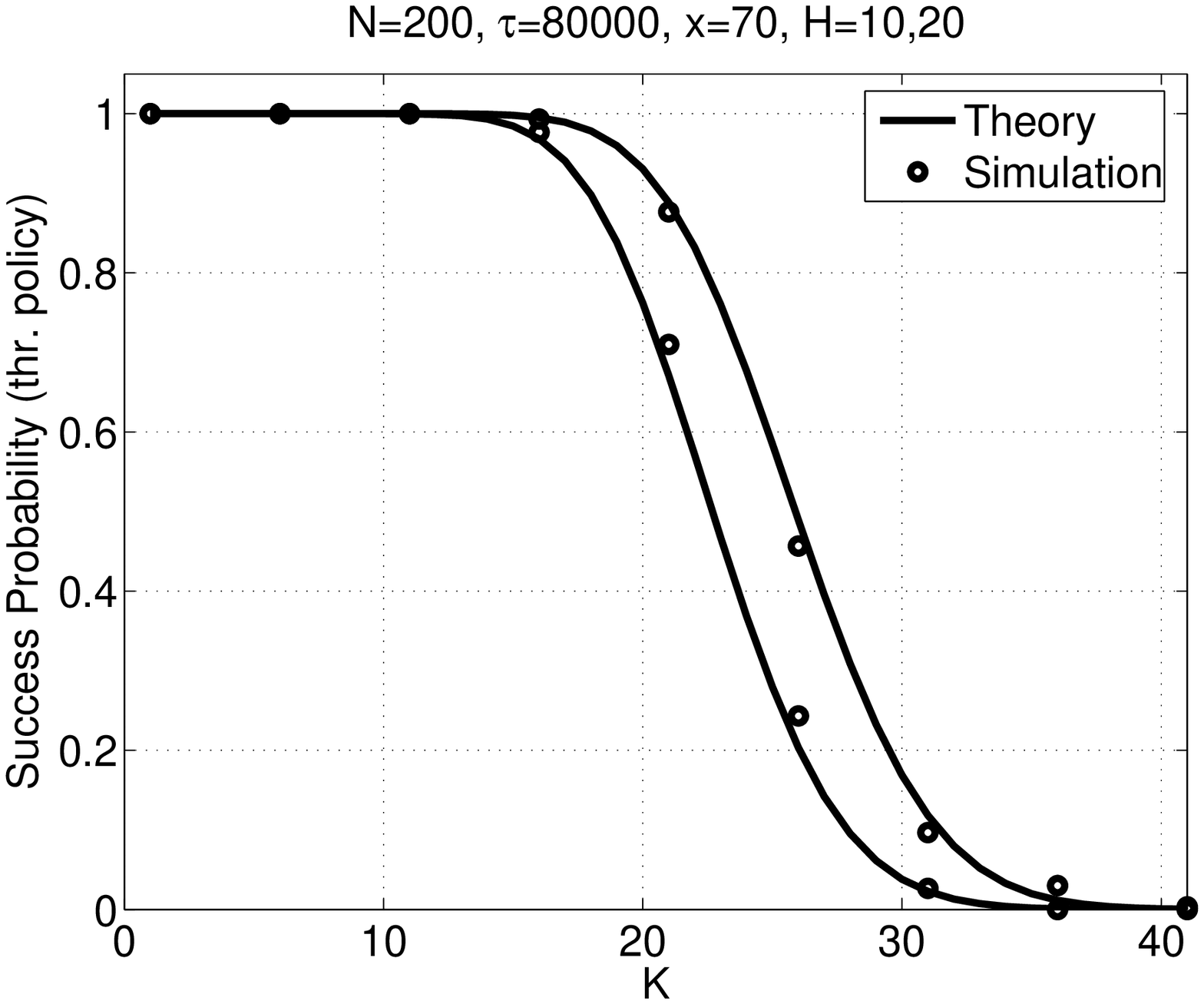}
\put(-35,80){\small $H=20$}\put(-38,77){\vector(-1,-1){10}}
\put(-105,35){\small $H=10$}\put(-70,45){\vector(1,1){10}}}\hskip5mm
\subfigure{\includegraphics [width=5cm] {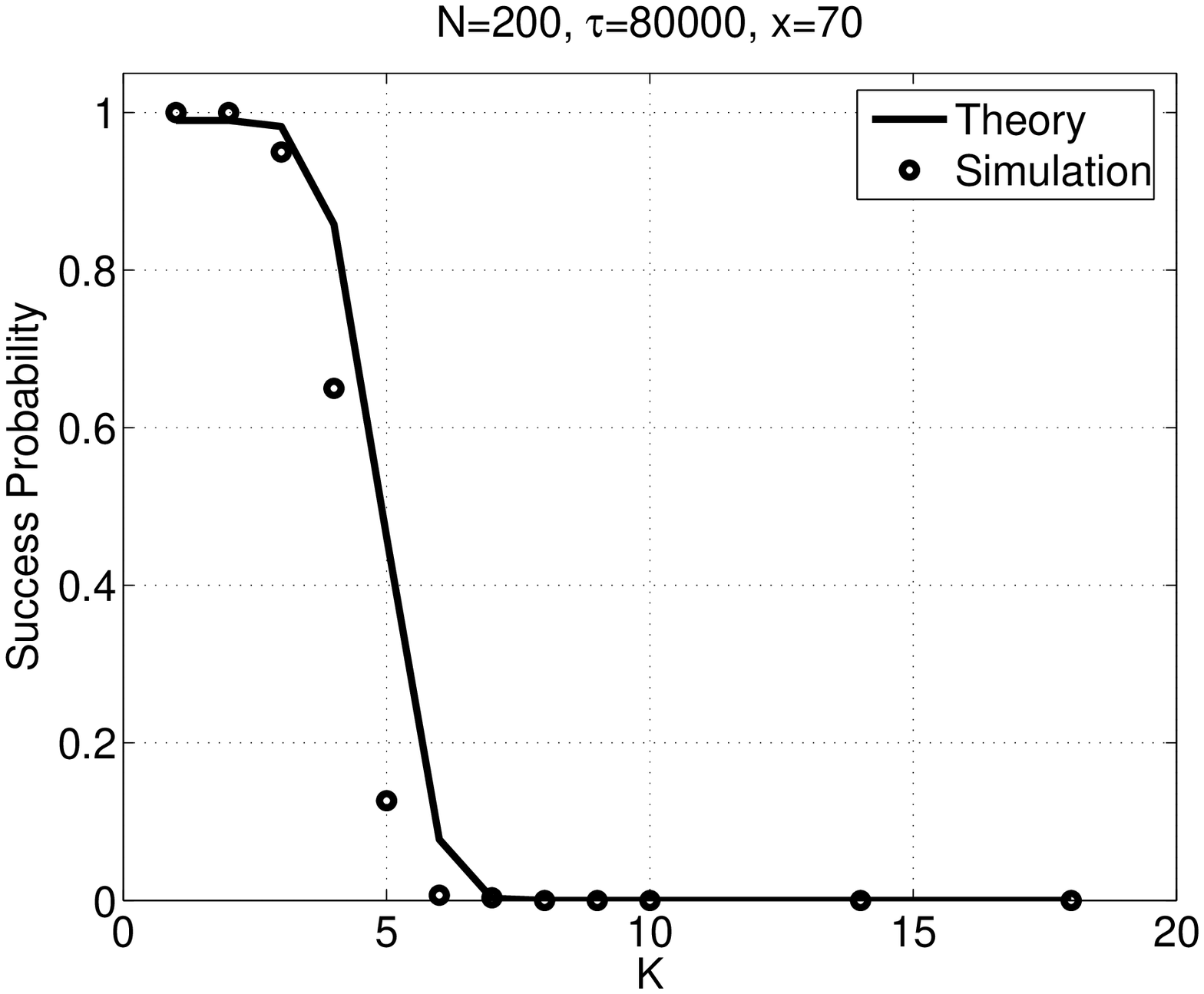}}
\put(-140,0){\put(-315,110){a)}\put(-160,110){b)}\put(-5,110){c)}}
\caption{a) Simulation results for the success probability of erasure coded messages, static policy, $H=10,20$, $N=200$, $\tau=80000$ s, $x=70$. b) Same as a) but under threshold policies c) Simulation results for the success probability of fountain coded messages, static policy, $N=200$, $\tau=80000$ s, $x=70$, $\delta=0.02$.}\label{fig:fig3}
\end{figure*}

\subsection{Comparison with static policies}

Here we would like to see what is the relative performance of static and threshold policies, both for fountain codes 
and erasure codes. In Fig.~\ref{fig:fig2}a) and Fig.~\ref{fig:fig2}b) we reported on the case a 
bound on energy exists $x=70$; in such case, $\sigma(z)=1000$ s. As concerns erasure codes, when the number of frames is
high ($30$) the usage of a large number of redundant frames proves much more effective compared to static policies. 
Conversely, a for a lower number of frames, the advantage of threshold policies is less marked. 

In the case of fountain codes, threshold policies are again more efficient than static policies, and the effect 
is more relevant for large values of the time constraint $\tau$. We recall that we refer to {\em optimal} static 
policies.

\section{Numerical Validation}\label{sec:simulations}

In this section we provide a numerical validation of the model. Our experiments are trace based; message delivery is 
simulated by a Matlab$^{\mbox{\scriptsize \textregistered}}$ script receiving as input pre-recorded contact traces. 

\subsection*{Synthetic Mobility} We considered first a Random Waypoint (RWP) mobility model \cite{camp02wcmc}. 
We registered a contact trace using Omnet++ with $N=200$ nodes moving on a squared playground 
of side $5$ Kms. The communication range is $R=15$ m, the mobile speed is $v=5$ m/s and the system starts in 
steady-state conditions in order to avoid transient effects \cite{boudec:info05}. The time limit is set to $\tau=80000$ s, 
which corresponds roughly to $1$ day operations, and the constraint on the maximum 
number of copies is $x=70$. With this first set of measurements, we want to check the fit of the model for the erasure codes 
and fountain codes. In the case of erasure codes, we fixed $H=10$ and $20$ and increased the number of message frames $K$. 
We selected at random pairs of source and destination nodes and registered the sample probability that the message is received 
at the destination by time $\tau$. As seen in Fig.~\ref{fig:fig3}a) the fit with the model is rather tight and an abrupt transition 
from high success probability to zero is visible. Also, in Fig.~\ref{fig:fig3}b) we reported on the results obtained in case of 
threshold policies; the fit is similar to what obtained for static policies, confirming the gain of performance with respect to 
static policies. 

We repeated the same experiment in the case of fountain codes, as reported in Fig.~\ref{fig:fig3}c) for static policies; in this case 
the code specific parameter is $\delta=0.02$ and again we increased $K$. Even in this case we see that the threshold effect predicted by 
the model is apparent. 

% \begin{figure}[h]
% \centering
% \includegraphics [width=6cm] {./FIG/simu_rwp_erasure.eps}
% \put(-60,100){$H=20$}\put(-63,97){\vector(-1,-1){30}}
% \put(-190,35){$H=10$}\put(-155,45){\vector(1,1){30}}
% \caption{Simulation results for the success probability of erasure coded messages, static policy, $H=10,20$, $N=200$, $\tau=80000$ s, $x=70000$}\label{fig:simu_rwp_erasure}
% 
% \end{figure}
% \begin{figure}[h]
% \centering
% \includegraphics [width=6cm] {./FIG/simu_rwp_fountain.eps}
% \caption{Simulation results for the success probability of fountain coded messages, static policy, $N=200$, $\tau=80000$ s, $x=70000$}\label{fig:simu_rwp_fountain}
% \end{figure}

\subsection*{Real World traces} 

Our model captures the behavior of a sparse mobile ad hoc network under some assumptions: the most stringent
is the uniformity and the stationarity of intermeeting intensities. We would like now to understand what 
is the impact of non-uniform and non-stationary encounter patterns. We considered two sets of experimental 
contact traces:

{\em  Haggle}:  in \cite{chaintreau07} and related works, the authors report extensive experimentation conducted in 
order to trace the meeting pattern of mobile users. A version of iMotes, equipped with a Bluetooth radio interface, 
was distributed to a number of people, each device collecting the time epoch of meetings with other Bluetooth devices. 
In the case of Haggle traces, due to the presence of several spurious contacts with erratic Bluetooth devices, 
we restricted the contacts to a subset having experienced at least $50$ contacts, resulting in $19$ active nodes.

{\em CN}: the CN dataset has been obtained by monitoring $21$ 
employee within Create-Net and working on different floors of the same building during a $4$-week period. 
Employees volunteered to carry a mobile running a Java application relying on Bluetooth connectivity. 
The application periodically triggers (every $60$ seconds) a Bluetooth node discovery; detected nodes are
recorded via their Bluetooth address, together with the current timestamp on the device storage for a later processing. 

\begin{figure*}[t]
\centering 
\subfigure{\includegraphics [width=5cm] {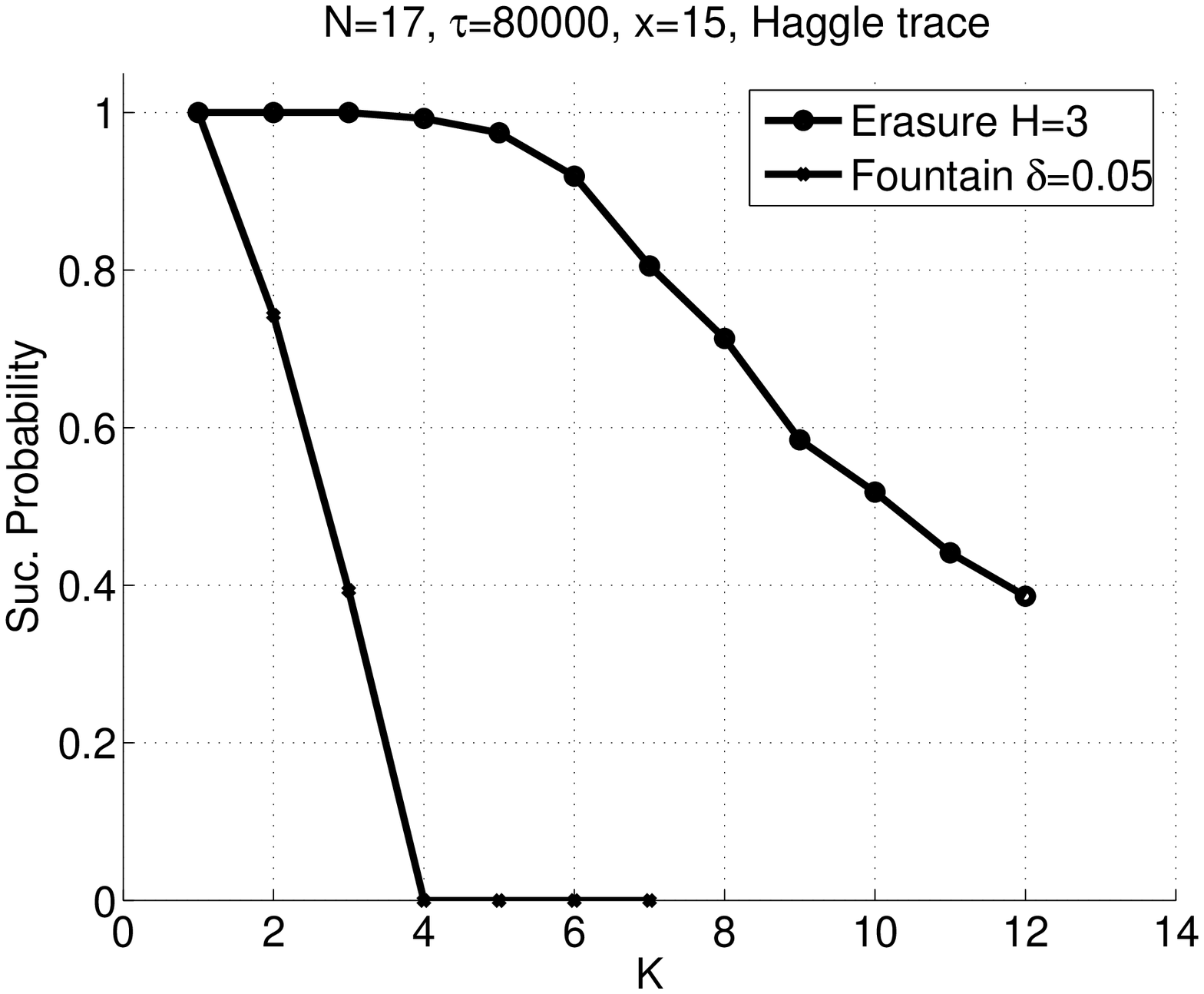}}\hskip10mm
\subfigure{\includegraphics [width=5cm] {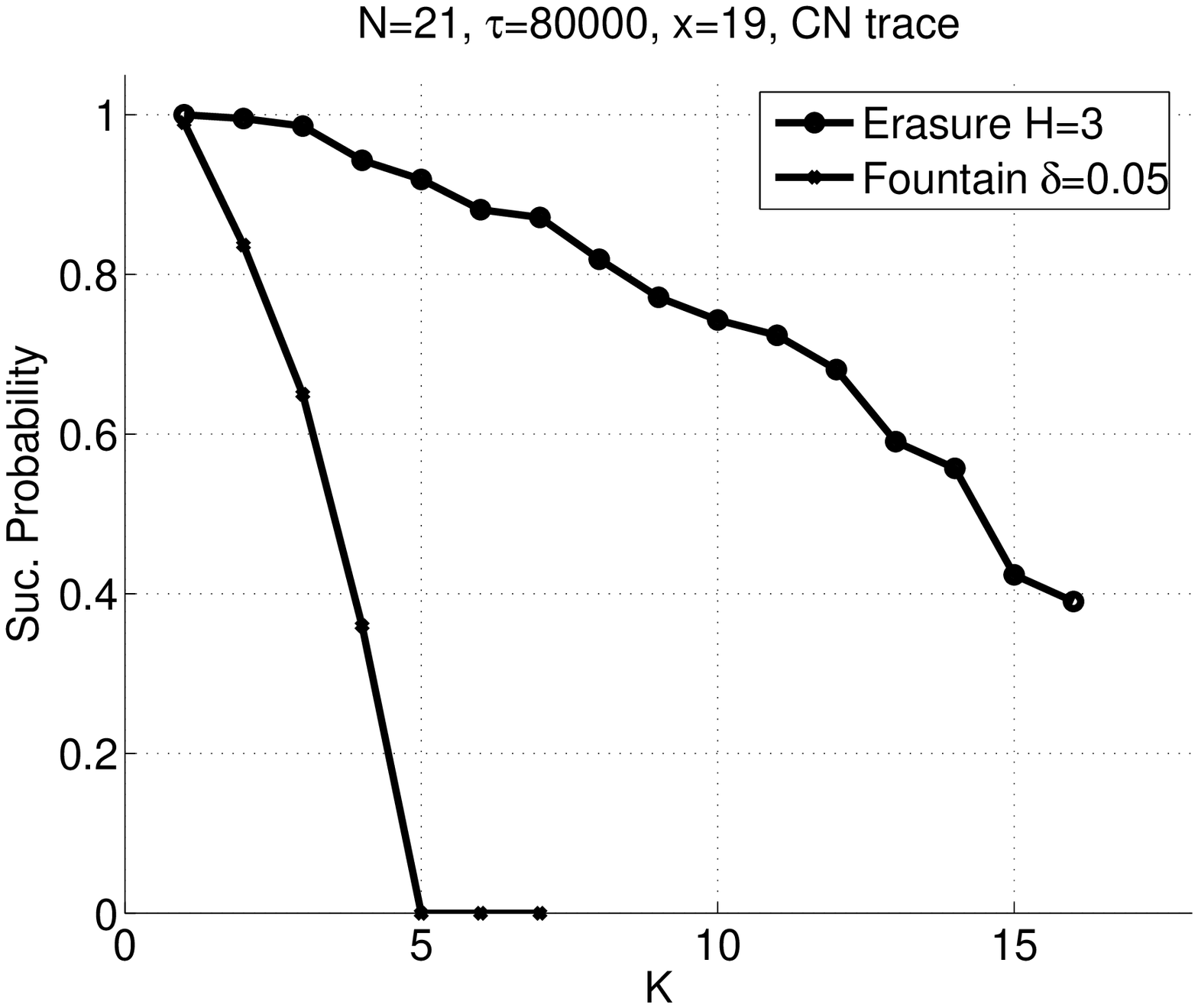}}
\put(-315,110){a)}\put(-142,110){b)}
\caption{a) Simulation results for the success probability of static policies, Haggle trace, 
$N=17$, $\tau=80000$ s, $x=17$ and $\delta=0.05$ b) Simulation results for the success probability of static policy, CN trace, 
$N=21$, $\tau=80000$ s, $x=17$ and $\delta=0.05$.}\label{fig:fig4}
\end{figure*}

Fig~\ref{fig:fig4} a) and b) depict the results of experiments performed with these data sets. A major impact
is played by the non-stationarity of traces. This is mainly due to ``holes'' appearing 
in the trace which impose unavoidable cutoff effects on the success probability. In 
particular, in the case of fountain coding, for $K>1$, i.e., when the original file 
is actually fragmented, the performance is rather poor. This is due to the increase with 
$K$ of the number of frames $M$ required for decoding, exacerbated by the reduced size 
of the network. For example, for $K=3$ and $\delta=0.05$ it holds $M=13$, i.e. $M$ 
is close to $N$ for the Haggle trace and larger than $N/2$ for the CN trace; this 
means that, in order to deliver the message, basically almost all nodes should 
 meet the destination and deliver a frame within $\tau$.

With erasure codes, both traces show the characteristic cutoff of performances at the increase 
of $K$. The decay, though, depends on the trace considered; in particular, a close 
look to the two data sets showed that the Haggle trace has higher average inter-meeting intensity 
compared to the CN trace over the considered interval; the shape seen in Fig~\ref{fig:fig4}a) 
resembles more closely the theoretical sigmoid cutoff predicted by the model.

\section{Conclusions and Remarks}

We have considered in this paper the tradeoff between energy and
probability of successful delivery in presence of finite duration of contacts 
or limited storage capacity at a node: it can store only part (a frame) 
of a file that is to be transferred. To improve performance we 
considered the generation of erasure codes at the source which allows 
the DTN to gain in spatial storage diversity. Both fixed erasure codes 
(Reed Solomon type codes) as well as rateless fountain codes have been studied.

Fountain codes can be viewed as special case of general network coding such as those
studied in \cite{FSCB,FBW,LLL}. We note however that in order to go
beyond fountain codes (in which coding is done at the source) and
consider general network coding (where coding is also done in the relay nodes),
one needs not only change the coding approach but
one should allow storage of several frames at each relay node.
%+-+ Added by Francesco
Also, interesting hints come from the presence of non-stationary
patterns arising in real world traces. New tradeoff issues arise which 
we shall study in future work.

%%%%%%%%%%%%%%%%%%%%%%%%%%%%%%
% References
%%%%%%%%%%%%%%%%%%%%%%%%%%%%%%

\bibliographystyle{IEEEtran}

%\bibliography{kbib}

\begin{thebibliography}{10}
\providecommand{\url}[1]{#1}
\csname url@rmstyle\endcsname
\providecommand{\newblock}{\relax}
\providecommand{\bibinfo}[2]{#2}
\providecommand\BIBentrySTDinterwordspacing{\spaceskip=0pt\relax}
\providecommand\BIBentryALTinterwordstretchfactor{4}
\providecommand\BIBentryALTinterwordspacing{\spaceskip=\fontdimen2\font plus
\BIBentryALTinterwordstretchfactor\fontdimen3\font minus
  \fontdimen4\font\relax}
\providecommand\BIBforeignlanguage[2]{{%
\expandafter\ifx\csname l@#1\endcsname\relax
\typeout{** WARNING: IEEEtran.bst: No hyphenation pattern has been}%
\typeout{** loaded for the language `#1'. Using the pattern for}%
\typeout{** the default language instead.}%
\else
\language=\csname l@#1\endcsname
\fi
#2}}

\bibitem{burleigh_commag03}
S.~Burleigh, L.~Torgerson, K.~Fall, V.~Cerf, B.~Durst, K.~Scott, and H.~Weiss,
  ``Delay-tolerant networking: an approach to interplanetary {I}nternet,''
  \emph{IEEE Communications Magazine}, vol.~41, pp. 128--136, June 2003.

\bibitem{chaintreau07}
A.~Chaintreau, P.~Hui, J.~Crowcroft, C.~Diot, R.~Gass, and J.~Scott, ``Impact
  of human mobility on opportunistic forwarding algorithms,'' \emph{IEEE
  Transactions on Mobile Computing}, vol.~6, pp. 606--620, 2007.

\bibitem{spy_ton08}
T.~Spyropoulos, K.~Psounis, and C.~Raghavendra, ``Efficient routing in
  intermittently connected mobile networks: the multi-copy case,''
  \emph{ACM/IEEE Transactions on Networking}, vol.~16, pp. 77--90, February
  2008.

\bibitem{tse_mobility02}
M.~Gr\"ossglauser and D.~Tse, ``Mobility increases the capacity of ad hoc
  wireless networks,'' \emph{IEEE/ACM Transactions on Networking}, vol.~10, pp.
  477--486, August 2002.

\bibitem{WJMK}
Y.~Wang, S.~Jain, M.~Martonosi, and K.~Fall, ``Erasure-coding based routing for
  opportunistic networks,'' in \emph{Proc. of SIGCOMM workshop on
  Delay-tolerant networking (WDTN)}.\hskip 1em plus 0.5em minus 0.4em\relax
  Philadelphia, Pennsylvania, USA: ACM, August 26 2005, pp. 229--236.

\bibitem{JDPF}
S.~Jain, M.~Demmer, R.~Patra, and K.~Fall, ``Using redundancy to cope with
  failures in a delay tolerant network,'' \emph{SIGCOMM Comput. Commun. Rev.},
  vol.~35, no.~4, pp. 109--120, 2005.

\bibitem{FBW}
C.~Fragouli, J.-Y.~L. Boudec, and J.~Widmer, ``Network coding: an instant
  primer,'' \emph{SIGCOMM Comput. Commun. Rev.}, vol.~36, no.~1, pp. 63--68,
  2006.

\bibitem{LLL}
Y.~Lin, B.~Liang, and B.~Li, ``Performance modeling of network coding in
  epidemic routing,'' in \emph{Proc. of MobiSys workshop on Mobile
  opportunistic networking (MobiOpp)}.\hskip 1em plus 0.5em minus 0.4em\relax
  San Juan, Puerto Rico: ACM, June 11 2007, pp. 67--74.

\bibitem{WB}
J.~Widmer and J.-Y.~L. Boudec, ``Network coding for efficient communication in
  extreme networks,'' in \emph{Proc. of the ACM SIGCOMM workshop on
  Delay-tolerant networking (WDTN)}, Philadelphia, Pennsylvania, USA, August 26
  2005, pp. 284--291.
  
\bibitem{FSCB}
A.~E. Fawal, K.~Salamatian, D.~C.~Y. Sasson, and J.~L. Boudec, ``A framework
  for network coding in challenged wireless network,'' in \emph{Proc. of
  MobiSys}.\hskip 1em plus 0.5em minus 0.4em\relax Uppsala, Sweden: ACM, June
  19-22 2006.

\bibitem{GNK}
R.~Groenevelt and P.~Nain, ``Message delay in {MANETs},'' in \emph{Proc. of
  SIGMETRICS}.\hskip 1em plus 0.5em minus 0.4em\relax Banff, Canada: ACM, June
  6 2005, pp. 412--413, see also R. Groenevelt, Stochastic Models for Mobile Ad
  Hoc Networks. PhD thesis, University of Nice-Sophia Antipolis, April 2005.

\bibitem{karagiannis07}
T.~Karagiannis, J.-Y.~L. Boudec, and M.~Vojnovi\'{c}, ``Power law and
  exponential decay of inter contact times between mobile devices,'' in
  \emph{MobiCom}.\hskip 1em plus 0.5em minus 0.4em\relax Montr\'{e}al,
  Qu\'{e}bec, Canada: ACM, September 9--14 2007, pp. 183--194.

\bibitem{ABD}
E.~Altman, T.~{Ba\c sar}, and F.~{{De} Pellegrini}, ``Optimal monotone
  forwarding policies in delay tolerant mobile ad hoc networks,'' in
  \emph{Proc. of ACM/ICST Inter-Perf}.\hskip 1em plus 0.5em minus 0.4em\relax
  Athens, Greece: ACM, October 24 2008.

\bibitem{ZNKT}
X.~Zhang, G.~Neglia, J.~Kurose, and D.~Towsley, ``Performance modeling of
  epidemic routing,'' \emph{Elsevier Computer Networks}, vol.~51, pp.
  2867--2891, July 2007.

\bibitem{Bollobas}
B.~Bollob\'as, \emph{Random Graphs}.\hskip 1em plus 0.5em minus 0.4em\relax
  Cambridge University Press, 2001.

\bibitem{McKay}
D.~J. MacKay, \emph{Information Theory, Inference, and Learning
  Algorithms}.\hskip 1em plus 0.5em minus 0.4em\relax Cambridge, UK: Cambridge
  University Press, 2003.

\bibitem{SPR}
T.~Spyropoulos, K.~Psounis, and C.~S. Raghavendra, ``Spray and wait: an
  efficient routing scheme for intermittently connected mobile networks,'' in
  \emph{Proc. of SIGCOMM workshop on Delay-tolerant networking (WDTN)}.\hskip
  1em plus 0.5em minus 0.4em\relax Philadelphia, Pennsylvania, USA: ACM, 2005.

\bibitem{camp02wcmc}
T.~Camp, J.~Boleng, and V.~Davies, ``A survey of mobility models for ad hoc
  network research,'' \emph{Wireless Communications \& Mobile Computing
  (WCMC)}, vol.~2, no.~5, pp. 483--502, August 2002.

\bibitem{boudec:info05}
J.-Y.~L. Boudec and M.~Vojnovic, ``Perfect simulation and stationarity of a
  class of mobility models,'' in \emph{Proc. of INFOCOM}.\hskip 1em plus 0.5em
  minus 0.4em\relax Miami, USA: IEEE, March 13--17 2005, pp. 183--194.

\bibitem{Feller}
W.~Feller, \emph{Introduction to Probability Theory and Its Applications},
  3rd~ed.\hskip 1em plus 0.5em minus 0.4em\relax New York: Wiley, 1971.

\end{thebibliography}

\section{Appendix} 

\subsection{Proof of Theorem \ref{thm3p1}.}
\begin{IEEEproof}The case $p=0$ is trivial so we assume below $p>0$.
We have
\begin{eqnarray}
\nonumber
P_s^*(\tau)&=&\max_{{\bf p}: \sum_i p_i = p } P_s(\tau)
=\max_{{\bf p}: \sum_i p_i = p }
\prod_{i=1}^{K} Z(p_i)
\end{eqnarray}
${\bf p^*}$ is optimal if and only if it maximizes
$ g({\bf p }) := \sum_{i=1}^{K} \log ( Z(p_i)) $ s.t.
$ \sum_i p_i = p $, $p_i \geq 0 $.
Note that $L(\tau)$ is non-positive.
$\log Z$ turns out to be a concave function. It then follows from
Jensen's inequality that
\begin{eqnarray}\label{log}
g({\bf p }) &=& \frac{1}{K} \sum_{i=1}^{K} \log ( Z(p_i))\nonumber \\
&\leq& \log \Big( Z\Big ( \frac{1}{K} \sum_{i=1}^{K} p_i \Big ) \Big)= \log \Big( Z\Big( \frac{p}{K} \Big ) \Big)
\end{eqnarray}
where equality is attained for $p_i=p/K$. This concludes the proof.\end{IEEEproof}
%\endpf

\subsection{Proof of Lemma \ref{equal}}
\begin{IEEEproof}Let $A(K,H)$ be the set of subsets $h \subset \{ 1,..., K+H \}$
that contain at least $K$ elements.
For example, $\{1,2,...,K\} \in A(K,H)$.
Fix $p_i$ such that $\sum_{i=1}^{K+H} p_i = p $.
Then the probability of successful delivery by time $\tau$
is given by
\[
P_s ( \tau,K,H ) = \sum_{ h \in A(H,K) } \prod_{i\in h} Z(p_i)
\]
For any $i $ and $j$ in $\{1,...,K+H\}$ we can write
\[
P_s(\tau,K,H) = Z( p_i ) Z ( p_j ) g_1
+ ( Z(p_1 ) + Z(p_2 ) ) g_2 + g_3
\]
where $g_1$, $g_2$ and $g_3$ are nonnegative
functions of $\{ Z(p_m) , m\not=i, m\not=j \}$.
For example,
\[
g_1 =  \sum_{ h \in A_{\{i,j\}} (H,K) } \prod_{
\stackrel{ m\in h }{ m\not=i },\;{ m\not=j}} Z(p_m)
\]
where $A_v(K,H)$ is the set of subsets $h \subset \{ 1,..., K+H \}$
that contain at least $K$ elements and such that $v\subset h$.

Now consider maximizing $P_s(\tau,K,H)$ over $p_i$ and $p_j$
that satisfy $p' = p_i + p_j $ for a given $p' \leq p$.
Since $Z(\cdot)$ is strictly concave, it follows by Jensen's inequality
that $Z(p_i)+Z(p_j)$ has a unique maximum at $p_i = p_j = p'/2$.
This is also the unique maximum of
the product $Z(p_i)Z(p_j)$ (using the same argument as in eq. (\ref{log})),
and hence of $P_s(\tau,K,H)$.

Since this holds for any $i$ and $j$ and for any $p' \leq p$,
this implies the Lemma.\end{IEEEproof}
%\endpf

\subsection{Proof of Proposition \ref{phase1}}
\begin{IEEEproof}From Thm.~\ref{thm:main_theorem},
\[
P_s^*(\tau,K,H)=P\{S_{H+K,\widehat p} \leq H+K \}-P\{S_{H+K,\widehat p}< K\},
\]
where $\widehat p=1-\exp \Big (\frac{L(p^*,\tau)p^*}{H+K}\Big)$. 

We can rewrite $-L( \tau , p^* )p^*=N\cdot \Gamma_0^{(0)} $, where we obtain
\[
\Gamma_0^{(N)}=-\frac {1}{p^*}  \Big ( 1- \lambda p^* \tau -
e^{-\lambda p^* \tau} \Big )=\lambda\tau 
\Big( 1+\frac{\frac{x}{N-z}}{\log(1-\frac{x}{N-z})} \Big)
\]
and we define
\[
\Gamma_0 := \lim_{N \to \infty} \Gamma_0^{(N)} = \lambda\tau 
\Big( 1+\frac{\widehat{x}}{\log(1-\widehat{x})} \Big)
\]

Notice also that, 
\begin{eqnarray}
E[S_{H+K,\widehat p}]&=&(H+K)\widehat p = (H+K)\Big(1-e^{-\frac {N\Gamma_0^{(N)}}{H+K}}\Big)\nonumber\\
                     &=&(H+K)\Big (\frac {N\Gamma_0^{(N)}}{H+K} + o\big (\frac{1}{H+K}\big )\Big)\nonumber
\end{eqnarray}
from which it follows that 
\begin{eqnarray}
\lim_{N\rightarrow \infty}\frac{E[S_{H+K,\widehat p}]}N&=&\lim_{N \to \infty} (H+K)\widehat p/N = \Gamma_0 \nonumber
\end{eqnarray}

Notice that the binomial r.v. $S_{H+K,\widehat p}$ can be interpreted as the sum of $H+K$ independent binary random 
variables $i_n$ with mean $\widehat p$ and variance $\widehat p(1-\widehat p)$. Also, the series of the normalized variances 
 $\sigma^2_{i_n}=\frac{1}{n^2}\widehat p(1-\widehat p)\leq \frac{1}{n^2}$ is finite. Thus, the Strong Law of Large Numbers \cite{Feller}
ensures that $ \lim_{N \to \infty} {S_{H+K,\widehat p}}/{N} = \Gamma_0 $ P-a.s.

Now we can derive  
\[
\lim_{N \to \infty} P\{S_{H+K,\widehat p}\leq H+K \}
=\begin{cases}
0, & \mbox{if} \quad \widehat K + \widehat H \leq \Gamma_0 \\
 1, & \mbox{if} \quad \widehat K > \Gamma_0 
\end{cases}
\]
and, in a similar way, 
\[
\lim_{N \to \infty} P\{S_{H+K,\widehat p}< K \}
=\begin{cases}
0, & \mbox{if} \quad \widehat K < \Gamma_0 \\
 1, & \mbox{if} \quad \widehat K \geq \Gamma_0 
\end{cases}
\]
Finally, by enumeration, 
\[
\lim_{N\to \infty} P_s(\tau,K,H)= 
\begin{cases}
0 & \mbox{if} \ \widehat K+\widehat H  \leq \Gamma_0\\[2mm]
0 & \displaystyle
 \mbox{if} \  \widehat K+\widehat H > \Gamma_0
   \ \mbox{and} \ \widehat K \geq \Gamma_0\\[2mm]
1 & \mbox{if} \
\displaystyle
\widehat K+\widehat H>\Gamma_0 \ 
   \mbox{and} \ \widehat K < \Gamma_0
\end{cases}
\]
This concludes the proof.\end{IEEEproof}
%\endpf
\end{document}